%% file: bare_jrnl_new_sample4.tex
\documentclass[lettersize,journal]{IEEEtran}
\usepackage{amsmath,amsfonts}
\usepackage{array}
\usepackage[caption=false,font=normalsize,labelfont=sf,textfont=sf]{subfig}
\usepackage{textcomp}
\usepackage{stfloats}
\usepackage{url}
\usepackage{verbatim}
\usepackage{graphicx}
\usepackage{cite}
\usepackage{booktabs}
\usepackage{algorithm}
\usepackage{algpseudocode}

\begin{document}

\title{6Rover: Leveraging Reinforcement Learning-based Address Pattern Mining Approach for Discovering Active Targets in IPv6 Unseeded Space}

\author{Zhichao Zhang, Zhaoxin Zhang, Yanan Cheng, Ning Li
\thanks{This work was supported in part by a grant from NSFC Grant no. 62101159, NSF of Shandong Grant no. ZR2021MF055, and also the Research Grants Council of Hong Kong under the Areas of Excellence scheme grant AoE/E-601/22-R. \textit{(Corresponding authors: Zhaoxin Zhang, Yanan Cheng)}}
\thanks{Zhichao Zhang is with the Faculty of Computing of Harbin Institute of Technology, Harbin, Heilongjiang 150001, China. E-mail: 22b303010@stu.hit.
edu.cn. }
\thanks{Zhaoxin Zhang, Yanan Cheng, and Ning Li are with the Harbin
Institute of Technology, Harbin, Heilongjiang 150001, China. E-mail: \{heart, chengyn, li.
ning\}@hit.edu.cn.}}



\maketitle

\begin{abstract}
The discovery of active IPv6 addresses represents a pivotal challenge in IPv6 network survey, as it is a prerequisite for downstream tasks such as network topology measurements and security analysis. With the rapid spread of IPv6 networks in recent years, many researchers have focused on improving the hit rate, efficiency, and coverage of IPv6 scanning methods, resulting in considerable advancements. However, existing approaches remain heavily dependent on seed addresses, thereby limiting their effectiveness in unseeded prefixes. 
Consequently, this paper proposes 6Rover, a reinforcement learning-based model for active address discovery in unseeded environments. To overcome the reliance on seeded addresses, 6Rover constructs patterns with higher generality that reflects the actual address allocation strategies of network administrators, thereby avoiding biased transfers of patterns from seeded to unseeded prefixes. After that, 6Rover employs a multi-armed bandit model to optimize the probing resource allocation when applying patterns to unseeded spaces. It models the challenge of discovering optimal patterns in unseeded spaces as an exploration-exploitation dilemma, and progressively uncover the potential patterns applied in unseeded spaces, leading to the efficient discovery of active addresses without seed address as the prior knowledge.
Experiments on large-scale unseeded datasets show that 6Rover has a higher hit rate than existing methods in the absence of any seed addresses as prior knowledge. In real network environments, 6Rover achieved a 5\% - 8\% hit rate in seedless spaces with 100 million budget scale, representing an approximate 200\% improvement over the existing state-of-the-art methods.
\end{abstract}


\section{Introduction}
\IEEEPARstart{A}{s}  the cornerstone of the next-generation networks, IPv6 has witnessed widespread deployment on a global scale. According to Google's statistics\cite{google_v6_stats}, the proportion of users accessing Google via IPv6 has consistently increased, reaching 45.17\% in October 2023, approximately 3.7 times higher than that of five years ago.  In the context of the ongoing rise in IPv6 deployment rate, there is substantial demand for IPv6 Internet-scale network surveys provide a better understanding of the real-world dynamics of IPv6 networks, including network topology measurements\cite{10.1145/3278532.3278559,JIA2019106947}, geolocation \cite{6search,gws-geo,hgl-geo,ipvseeyou}, Internet services evaluation\cite{durumeric2013zmap, izhikevich2021lzr}, and security analysis\cite{li2021fast,li2020towards,10.1145/3544912.3544915,rye2021follow}. However, unlike IPv4, due to the vast address space of the IPv6 network, it is infeasible to perform exhaustive scans of all addresses\cite{6gen}. Consequently, the implementation of active address generation stands as a prerequisite for achieving IPv6 Internet-scale network surveys. 

In 2012, when Google's IPv6 adoption rate was only 0.7\%, researchers like Barnes had already started exploring methods for discovering active addresses in IPv6\cite{barnes2012mapping}. Over the past five years, as this figure has approached nearly 50\%, discovering active addresses in the extensive IPv6 space has emerged as a critical issue in network measurements. Thus, during this period, a significant number of researchers have attempted to discover as many active IPv6 addresses as possible, mainly through passive collection \cite{pam2017-something-from-nothing-there, DBLP:conf/sp/BorgolteHFV18, gasser2018clusters, gasser2022,10.1007/978-3-319-76481-8_10 } and active probing \cite{6gen,6forest,6graph,6gan,6veclm,6hit,6gcvae,6tree,6scan}. 

In existing approaches, active probing methods exhibited remarkable improvements in probing efficiency with a hit rate of 6\% in 2017 using 6Gen \cite{6gen} to 57\% in 2022 using Addrminer \cite{addrminer} under a similar probing budget scale. However, those algorithms heavily rely on the quality and quantity of known seed addresses as indispensable prior knowledge. Therefore, these methods are ineffective in discovering addresses in unseeded spaces.

\begin{figure*}[ht]
    \centering
    \includegraphics[width=0.75\textwidth]{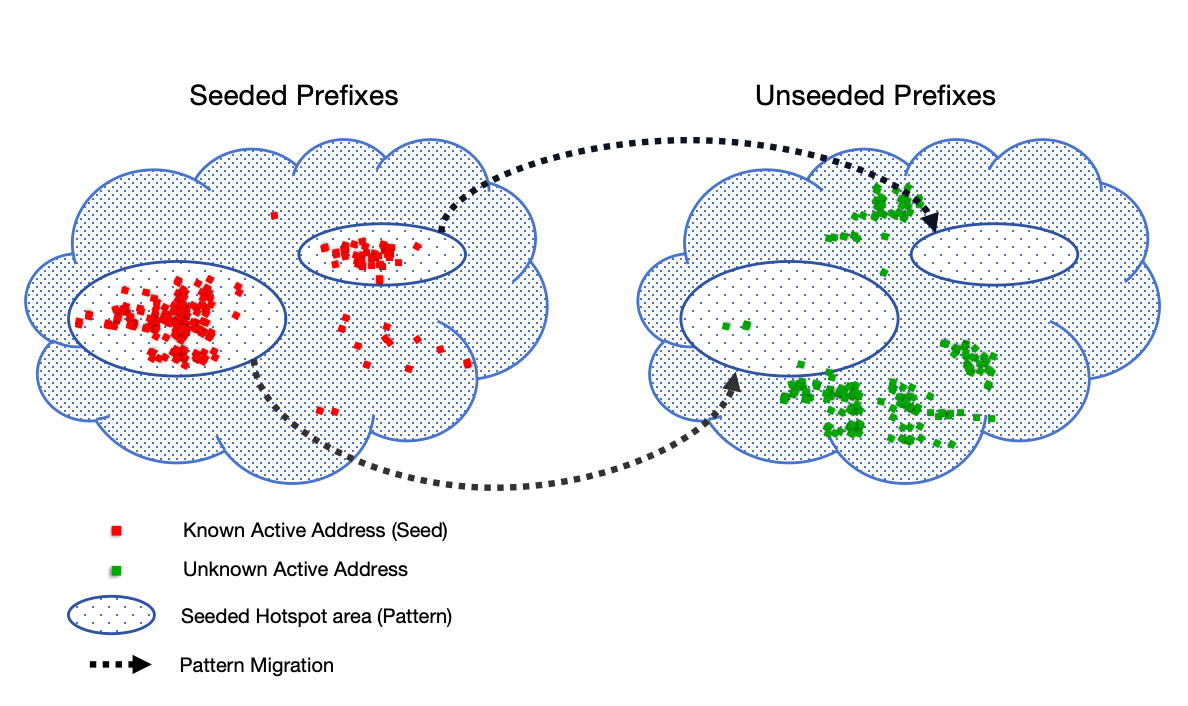}
    \caption{\textbf{The diagram of discovering active addresses in unseeded prefix by pattern migration.} It can be observed that due to the different distributions of active addresses between seeded prefixes and unseeded prefixes, the direct migration of address patterns results in a very low hit rate.}
    \label{fig:pattern_migration}
\end{figure*}

 Obviously, discovering active addresses in unseeded spaces is crucial because existing methods fail to provide a comprehensive and unbiased view of the IPv6 network. However, although more researchers are concentrating on enhancing the coverage of active IPv6 addresses, the effectiveness in detecting active addresses within unseeded spaces is still limited. 
 
 There are two existing feasible strategies for discovering active addresses in unseeded space. The first strategy is directly collecting addresses in unseeded spaces to create a hitlist with higher coverage, thereby transforming unseeded spaces into seeded spaces. This includes both empirical methods\cite{6scan, hmap6} and hitlist construction techniques \cite{gasser2018clusters, gasser2022, DBLP:conf/sigcomm/RyeL23}.  Although such methods pragmatically increase knowledge of unseeded spaces, the addresses collected from these areas heavily depend on the collection method, leading to potential biases. For example, heuristic methods \cite{6scan, hmap6} can only acquire addresses that conform to specific rules. Meanwhile, Since addresses obtained through hitlist construction methods \cite{gasser2018clusters , gasser2022, DBLP:conf/sigcomm/RyeL23} depend on input sources such as DNS resolution, traceroute, or passive sources, the types and quantities of addresses collected are limited, and those methods are less effective in actively discovering addresses in designated unseeded spaces.

The second strategy involves the migration of address patterns from seeded to unseeded spaces. As shown in Figure \ref{fig:pattern_migration}, the term 'address patterns' represents a clustering of seed address distributions. These methods correlate seeded prefixes with unseeded ones based on specific characteristics, facilitating the migration of patterns and enabling active scanning for potential active targets in unseeded spaces. For example, the state-of-the-art method Addrminer\cite{addrminer} employs WHOIS information of prefixes as the associative feature to migrate patterns. However, the practical address allocation strategies configured by network administrators do not clearly and widely manifest in the public prefix's attributes such as WHOIS fields and organization names. In most cases where the WHOIS information of prefixes are entirely unrelated, the algorithm resorts to a strategy of random matching from the common pattern library, leading to lower hit rates of merely 3\% under a small budget size(i.e., 3k). 


The primary challenge of discovering active addresses in unseeded prefixes lies in the absence of seed addresses as prior knowledge. This absence hinders the understanding of address patterns within unseeded space, including the distribution of active subnets and IID allocations, which is vital for active address prediction. Consequently, active probing algorithms risk wasting numerous probe packets in searching for employed address patterns in unseeded space, resulting in a very low hit rate. Moreover, without evidence of any active addresses (seed) in unseeded prefixes, algorithms struggle to ascertain the presence of discoverable addresses. Persistently probing a non-operational or privacy-intensive prefix leads to wasted probe packets without uncovering any active addresses.

In this work, we proposed 6Rover, a reinforcement learning-based address pattern mining algorithm focusing on the discovery of active IPv6 addresses in unseeded environments to solve the aforementioned challenges. We transform the challenge of discovering active addresses in unseeded prefixes in the problem of exploring and exploiting the potentially optimal address patterns. To represent the actual strategies of network administrators in address allocation, 6Rover converts the address patterns from seeded space into \textit{generic patterns}, concealing the specific hotspot distributions in seeded spaces to avoid the mistake of \textit{marking a boat to locate a lost sword}. After that, 6Rover employs an efficient parallel multi-armed bandit model to gradually identify the most effective generic patterns in unseeded prefixes. This strategy efficiently balances exploration with exploitation in a parallel architecture, leading to an improved overall hit rate in target generation. 

In order to demonstrate the validation of 6Rover in our real-world network, we tested the performance of discovering active targets within unseeded prefixes without any specific seed information. In contrast to the state-of-the-art methods \cite{addrminer, hmap6}, the real-world test results show that 6Rover exceeds existing baseline methods by more than 200\% in terms of hit rate, reaching 5\% - 8\% with a 100M budget scale.

In summary, the main contributions  of the paper are as follows:
\begin{itemize}
\item  We present 6Rover, a reinforcement learning-based IPv6 address pattern mining algorithm, which aims to discover active IPv6 addresses efficiently in unseeded environments. Without any prior knowledge of active address distribution in the targeted prefix, the algorithm can still infer potential patterns usage through dynamic scanning by reinforcement learning, generating addresses with higher hit rates. Consequently, this method enables researchers to conduct further security-related measurements and studies with limited seed address quantities and coverage.\\
\item  6Rover implements a pattern construction algorithm to establish a more generalized pattern, tailored for the task of target generation in unseeded prefixes. This method solved the pattern bias issues of existing address patterns, avoiding the inefficient emphasis on specific and limited hotspots of seeded prefixes, which leads to probe packet wastage. \\
\item  6Rover models the challenge of using the minimum number of probe packets to identify the optimal address pattern set from several potential generic patterns in unseeded prefixes as an exploration-exploitation problem in reinforcement learning. Utilizing the multi-armed bandit model, 6Rover dynamically adjusts pattern selection strategy based on real-time scanning feedback and assessment of the actual active address distributions, thereby identifying the best generic pattern set for unseeded prefixes with reduced probing costs.\\
\item Real network experimental results show that 6Rover achieves up to a 5\% to 8\% hit rate with a 100M budget, which is a more than 200\% improvement over the state-of-the-art methods. 
\end{itemize}

The remainder of this paper is organized as follows. In Section 2, we introduce the related work of active IPv6 address probing both in seed and unseeded space. In Section 3, we outline the preliminaries of this paper, including the definitions of fundamental concepts and problems. Section 4 presents the details of the 6Rover algorithm. We evaluate the real-world performance in Section 5. Finally, we conclude this paper in Section 6.

\section{Related Work}
Given the vastness and sparseness of the IPv6 address space, brute-force measurement approaches by scanning the complete address space are now considered entirely unfeasible\cite{rfc7707, gasser2018clusters} . Consequently, to discover active addresses in IPv6 networks via active scanning, many researchers have pursued intelligent scanning methods, aiming to enhance the quality, quantity, and efficiency of the IPv6 target generation. In this section, we review the related works of IPv6 address generation problem, dividing the existing work into the following three categories:
\subsection{Address Clustering-Based Approach }
This type of method takes a set of known IPv6 addresses as prior knowledge, examines the address patterns and clustering characteristics, and then strategically probes the hotspot areas from the clustered address results. These seed addresses serving as prior knowledge are typically collected from public resources or passive traffic. A notable contribution is Gasser's Hitlist\cite{gasser2018clusters, gasser2022, ipv6_hitlist}, which continually gathers active IPv6 addresses from sources such as DNS and routing data, maintaining an updated hitlist that forms an essential prerequisite for the operation of these algorithms. 

Specifically, Murdock et al. \cite{6gen} introduced 6Gen, initially proposing the concept of seed addresses and presenting a key hypothesis widely adopted in many later algorithms: address spaces with high-density seed addresses have greater potential for discovering more active addresses. 6Gen initially clusters similar addresses to construct high-density address spaces, then employs a greedy strategy to iteratively probe those dense areas, aiming at maximizing the density of active addresses with the least number of probe packets. 	Liu et al. introduced 6Tree \cite{6tree}, implementing a dynamic target generation method based on iteratively applying Divisive Hierarchical Clustering (DHC) to known seed addresses and recently detected addresses. Consequently, it enables scan direction adjustment in linear time complexity based on real-time scanning results. This method also integrates alias detection technology into the scanning module, preventing the wastage of probe packets in alias regions in real-time scanning. Hou et al. introduced 6Hit \cite{6hit}, which applies reinforcement learning to active address scanning in seeded spaces. Unlike 6Tree's dynamic scanning, 6Hit considers the hit rate from each probe as a reinforcement learning reward and assigns probe packets for next iteration scans based on the total revenue of each address space, effectively targeting high-value regions. 

At this time point, researchers have established a basic framework for detecting active addresses in seeded spaces, which involves initially constructing an address space pattern tree that reflects hotspot regions using the DHC algorithm, followed by strategically selecting the most appropriate scanning sequence or targeted regions. During this period, DET \cite{det}, 6Graph\cite{6graph}, and 6Forest\cite{6forest} were introduced, each attributing the low hit rates of current methods to the inappropriate space partition caused by the short-sighted splitting indicators in the DHC algorithm. To address this issue, DET and 6Forest have discarded the leftmost splitting indicator of the DHC space partition, introducing the minimum entropy splitting indicator and maximum-covering splitting indicators, respectively. Additionally, 6Graph has developed an unsupervised outlier elimination algorithm based on graph theory. All three methods have effectively improved the hit rate. Subsequently, after various algorithms achieved relatively high hit rates, Hou introduced 6Scan\cite{6scan}, which optimizes both the probing speed and hit rate. 6Scan, building upon 6Hit, incorporates Regional Identifier Encoding in the packet header to rapidly adjust scanning spaces. It enables asynchronous scanning, saving time in evaluating the feedback of each subspace, and thereby enhancing overall speed. In summary, these methods, based on the clustering results of known seed addresses, have achieved enhancements in both the hit rate and speed of active address discovery tasks. 

However, these methods are critically dependent on the quantity and quality of seed addresses, achieving higher hit rates primarily in high-density areas. Additionally, since the patterns extracted from seed addresses by these methods are only applicable within the prefixes of the seed addresses, these algorithms are entirely incapable of operating in prefixes without seeds, significantly limiting their applicability.

\subsection{Deep Learning-Based Approach}
Another method involves using deep learning models to address the IPv6 target generation problem. These methods use a large number of IPv6 addresses as input, attempting to extract semantic information from the addresses through deep learning models and generate target addresses based on natural language processing techniques. Cui et al. proposed 6GCVAE\cite{6gcvae}, 6VecLM\cite{6veclm}, and 6GAN\cite{6gan} to enhance the hit rate of the target generation algorithm. In these approaches, IPv6 address generation is reframed as a textual generation problem in the domain of natural language processing. These approaches model each nibble of an IPv6 address as a word and the complete address as a sentence, utilizing Variational Autoencoder\cite{kingma2022autoencoding}, Transformer\cite{NIPS2017_3f5ee243}, and Generative Adversarial Network\cite{10.1145/3422622} to solve text generation problems. 

To sum up, this type of method maps the nibbles of IPv6 addresses to a semantic space based on the known seed address distribution, capturing potential dependencies among the nibbles by hidden layers. Leveraging the learned seed address distribution and employing text generation models successfully in NLP, it can generate target addresses with similar distribution and achieve a relatively high hit rate. This strategy is fundamentally different from the aforementioned address clustering-based approach. 

However, the deep learning-based approach has two main drawbacks. Firstly, since this method is based on deep learning models, it lacks a certain degree of interpretability. It is difficult to ascertain whether the address generation model trained on seed addresses can generalize to unseeded space. Consequently, this approach also heavily relies on the quantity and quality of seed addresses. Secondly, the deep learning networks used in these methods have high computational power requirements and are very time-consuming to train on large datasets, hindering the suitability for large-scale scanning.

\subsection{Approach Working in Unseeded Prefixes}
The primary challenge addressed by such methods revolves around generating active probing targets within an unseeded prefix space while enhancing the coverage of active probing algorithms within all announced BGP prefixes. Firstly, Song et al. highlighted the challenge of low hitlist coverage within all announced BGP prefixes, and introduced Addrminer \cite{addrminer} to generate probing targets in unseeded prefix spaces. Addrminer utilizes a pattern migration strategy, clustering addresses in seeded prefixes using DHC to extract address patterns, and then selectively migrating these patterns to unseeded prefixes. In Addrminer, the basis for determining a pattern migration relationship between two prefixes is their Whois records. A pattern-sharing relationship is established when there's a high similarity in the word embedding vectors of the organization names of the two prefixes. 

Subsequently, Song introduced an enhanced version of Addrminer \cite{addrminer_2.0}, which identifies prefixes likely to share the same address allocation patterns with other prefixes by counting the occurrences of overlap in Whois fields, such as organization name, AS number, naming authority, description, email, phone, and administrator ID, between the non-seeded prefix and all other seeded prefixes. This method is capable of identifying several prefixes within the same Autonomous System (AS), as well as prefixes under various ASes belonging to the same organization. This set of prefixes, including both seeded and non-seeded, has a high probability of sharing similar address allocation patterns. However, this method faces a primary drawback. Beyond the aforementioned directly related cases, the similarity in prefix Whois information is insufficient to represent the relevance of address pattern allocation strategies, leading the algorithm to predominantly rely on random pattern migration; consequently, the algorithm achieves only a 3\% hit rate at a smaller budget scale (i.e. 10000 packets). 

Additionally, Hou et. al introduced HMap6 \cite{hmap6}, aiming at reducing reliance on seed addresses. HMap6 employs a heuristic method to collect addresses from all announced BGP prefixes, thereby transforming the problem of generating targets in non-seeded prefixes into seeded prefixes. It then uses a cluster-based approach to identify hotspot areas and generate target addresses. The key to generating addresses under unseeded prefixes in HMap6 lies in its heuristic method of address collection. This method systematically determines active address spaces nibble by nibble, exploring all 16 possibilities for the current prefix to form a subnet part, combined with the "::1" IID portion to construct a complete address. Finally, this method reduces the exploration of spaces containing inactive addresses and focuses on conducting further probes in the subnet spaces where active addresses are located. However, the success of the heuristic method relies on an assumption: if an active address exists, addresses formed by concatenating a shorter subnet from this address with the IID ('::1') are also active. Consequently, active addresses in unseeded prefixes that do not adhere to this assumption are not collected by the heuristic method, leading to bias on the target generation result.
\section{Preliminaries}
\subsection{Definitions}
\subsubsection{Unseeded Address Space}
A set of announced BGP prefixes with no known active address from the current hitlist. When announced BGP prefixes set is $G$, the set of active seeded addresses (hitlist) is $H$, and known active addresses are represented as $s$, the unseeded address space N has the following relations: 
\[
\forall s \in H, s \notin N \quad \text{and} \quad N \subset G
\]
\subsubsection{Address Patterns}
An address allocation pattern indicates the distribution of address sets under a specific prefix in terms of the IID and local subnet, typically comprising digits, ranges, lists, and wildcards. It is noteworthy that the address pattern mentioned here differs from the pattern in RFC 7707\cite{rfc7707}. In this context, the address pattern is derived from clustering seeded addresses using DHC, followed by the exclusion of outliers, resulting in a summarized address string used for identifying active addresses in seeded regions. 

\subsubsection{Generic Patterns}
\begin{figure}[t] 
    \centering 
    \includegraphics[width=\linewidth]{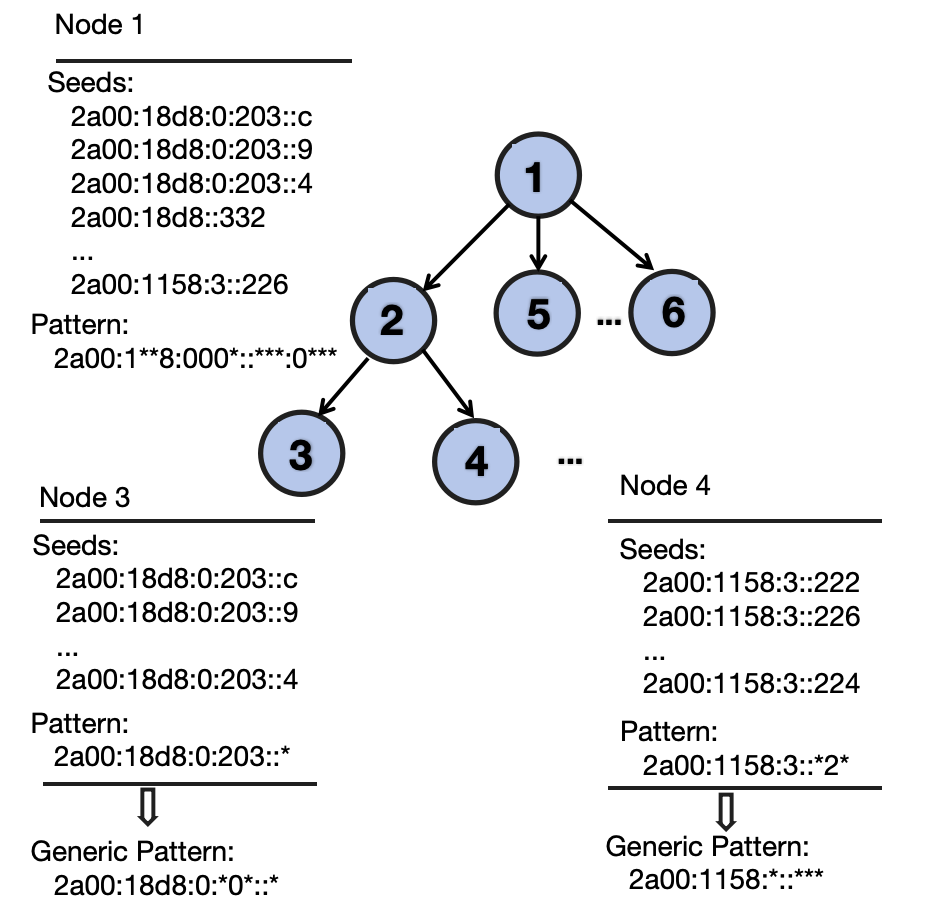} 
    \caption{The illustration of patterns and generic patterns} 
    \label{fig:generic_pattern_illu} 
\end{figure}

A group of address patterns consisting only of 0 and *, and employed by a minimum of two prefixes or Autonomous Systems. As shown in figure \ref{fig:generic_pattern_illu}, the generic pattern treats non-zero nibbles in the address pattern as wildcards, which, although expanding the scanning space, represents a more generalized address allocation strategy, making it suitable for pattern discovery in unseeded spaces.

\subsubsection{Non-aliased hit rate}
The non-aliased hit rate $d$ is calculated using the following formula where $c$ is the total probe packets number, $a$ is the number of alias addresses, and $p$ is the number of active addresses. In actual IPv6 probing, a high number of alias addresses can give an illusion of a high hit rate, but the actual probing data might be subpar. This becomes more significant in unseeded space probing, as mistaking aliases for correct addresses can lead to misguided probing and the expenditure of many packets on invalid areas. 
\[d=(p-a)/ c\]

\subsection{Problem Statement }
The main problem solved by 6Rover is the generation of non-alias addresses in unseeded spaces while maintaining a high hit rate. This challenge differs significantly from address discovery in seeded spaces, as it involves partitioning the entire IPv6 space into seeded and unseeded sections. The goal is to leverage address data from the seeded spaces to discover active addresses in the unseeded spaces. Given that $S_p$ is the set of seeded addresses, P and Q represent the sets of seeded and unseeded prefixes, respectively, b denotes the budget for probe packets, $A_q$ is the set of actual active addresses in unseeded prefixes, and $F_q$ is the set of alias addresses in the unseeded prefixes, the goal is to find an algorithm f that maximizes the metric defined as follows:

\begin{align}
    \max \quad & \frac{[f(S_p, b) - F_q]\cup A_q}{b} \nonumber\\
    \text{s.t.} \quad & S_p \in P \nonumber \\
                      & A_q \in Q \nonumber \\
                      & f(S_p, b) \in Q \nonumber \\
                      & Q\cup P = \emptyset \nonumber 
\end{align}
where the formula is constrained by the conditions that all seeded addresses $S_p$ belong to the seeded space P, active addresses $A_q$, and algorithm-generated addresses $f(S_p, b)$ are part of the unseeded space $Q$, with no overlap between the seeded space P and unseeded spaces $Q$.

\section{Design of 6Rover}
In this section, we will first outline the workflow of 6Rover, followed by a detailed explanation of its two main technical aspects: Generic Pattern Conversion and Exploration \& Exploitation of Generic Pattern.

\subsection{System Overview}
\begin{figure*}[ht]
    \centering
    \includegraphics[width=0.85\textwidth]{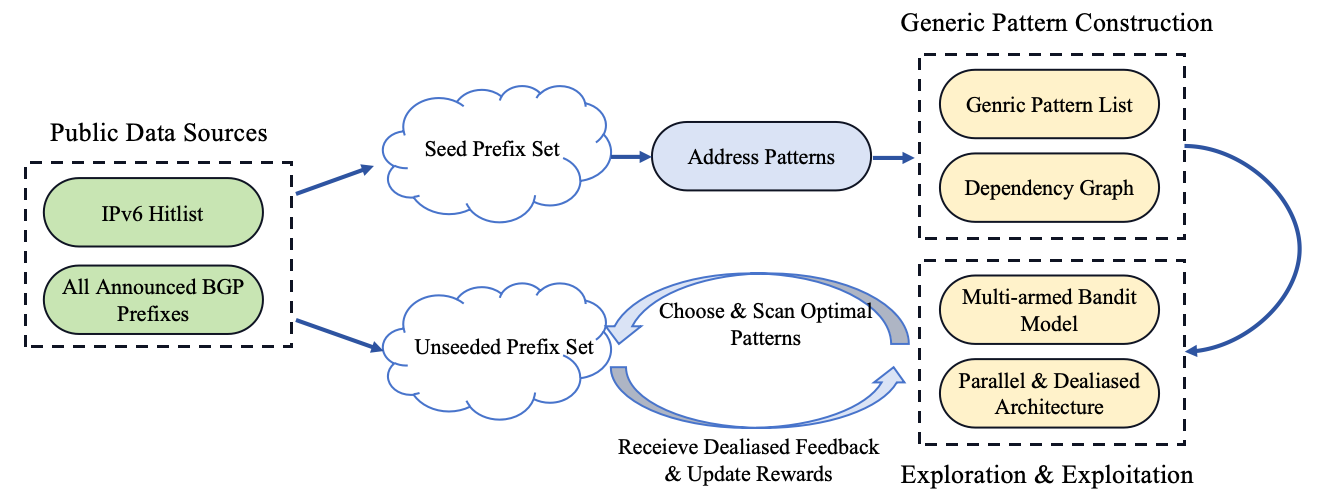}
    \caption{6Rover Overview}
    \label{fig:architecture}
\end{figure*}

Figure \ref{fig:architecture} illustrates the main workflow of 6Rover. The input of 6Rover are all announced prefixes list collected from RIPE\cite{RIPEstat}, and hitlist merged from the published address datasets including Gasser\cite{ipv6_hitlist, gasser2018clusters, gasser2022}, HMap6 \cite{hmap6} and Addrminer\cite{addrminer}. Firstly, it matches addresses from the Hitlist with all announced prefixes to find the unseeded prefix set as the target space for the following address generation. Additionally, all IPv6 addresses in the Hitlist are scanned for 7 days and only continuously active addresses are preserved. Using the alias prefix list provided by Gasser\cite{ipv6_hitlist, gasser2018clusters, gasser2022} and the de-aliasing method proposed by \cite{gasser2018clusters}, alias prefixes are also identified and removed to construct a clean collection of active seeded addresses. After that, we utilize the DHC algorithm to transform addresses into address patterns, and then convert these patterns into generic patterns and obtain generic pattern dependency through the generic pattern conversion module. Ultimately, 6Rover systematically traverses the prefixes in the unseeded prefix set, establishing a multi-armed bandit model for every unseeded prefix. In this model, various candidate generic patterns serve as arms, facilitating an address discovery process that optimally balances exploration and exploitation. The generic patterns are chosen in an ascending order of complexity for each probing cycle. Based on the dependency relationship of these patterns, increasingly complex and dependent generic patterns are identified for the next round of exploration as arms. In the end, 6Rover can discover the applied patterns in different unseeded prefixes, resulting in a hit rate that exceeds existing state-of-the-art techniques.

\subsection{Generic Pattern Extraction }
When discovering active addresses in seeded spaces, it is common to convert known active IPv6 addresses into address patterns\cite{6forest,6tree, 6graph, addrminer}. This helps to reduce the size of scanning space by focusing on the hotspot areas defined by these patterns, which potentially contain more active targets.

However, when probing unseeded spaces, existing methods still directly apply patterns constructed in seeded spaces to these unseeded spaces\cite{addrminer}, leading to a problem of \textit{pattern bias}. This bias is comparable to the sampling bias outlined by \cite{6hit}, which highlights a disparity between the distribution of seeded addresses and active addresses in seeded spaces. Likewise, pattern bias signifies the variance between address patterns extracted from seeded address distributions and the actual active address distribution in unseeded spaces. For instance, the pattern 0008:0000:0000:000f:2*** is summarized from several seed addresses in the seeded space. When it is directly applied to an unseeded prefix, there is a potential assumption that a similar hotspot area exists in the unseeded space with the 4th nibble as 8, the 6th as f, and the 17th as 2. However, the existence of this hotspot area lacks evidence from unseeded space, leading to potential misjudgments about the address distribution.

Consequently, 6Rover introduces \textit{Generic Pattern} specifically designed for unseeded space address discovery. While it is also based on addresses from seeded spaces, it is more generalized and abstract compared to current address pattern models, striving to represent the common logical principles used in IPv6 address distribution by network administrators.

\begin{figure}[ht] 
    \centering 
    \includegraphics[width=0.8 \linewidth]{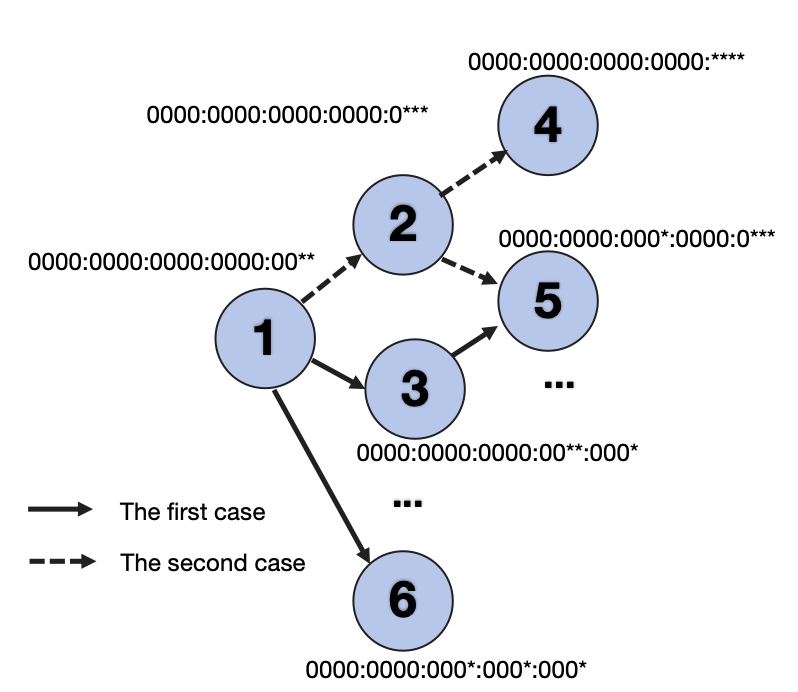} 
    \caption{The illustration of dependency graph of generic patterns. } 
    \label{fig:dependency} 
\end{figure}

Algorithm 1 and Algorithm 2 respectively illustrate the main process of Generic Patterns Construction and the associated Utility Functions. This algorithm takes a set of seeded addresses $C$ and a mapping of seed addresses to prefixes as inputs, and produces a Generic Pattern List $G$ and the Dependency Graph $D$ as outputs. Specifically, as illustrated in Figure \ref{fig:dependency}, the Dependency Graph $G$ is a graph representation of the inter-dependencies between Generic Patterns. In this graph, each vertex links to other related vertices with higher scanning space. Upon discovering an efficient pattern, this graph can be further utilized as a springboard to delve deeper into detecting other related and more complex patterns.

Firstly, from the first to the second line of Algorithm 1, 6Rover masks the first 48 bits of seed addresses and employs existing methods based on Divisive Hierarchical Clustering to construct address patterns. According to the RFC recommendations \cite{rfc6177}, we treat both the subnet ID (48th to 64th) and Interface ID (64th to 128th) as the address pattern segments because network administrators may manually assign these fields. Note that the function $PatternMining$ in the second line of the Algorithm can be the existing DHC-based pattern mining algorithm in seed space, such as 6Graph \cite{6graph}, 6Forest\cite{6forest} and DET\cite{det}, and we utilize DET \cite{det} in the following experiments.

Additionally, from line 3 to line 7 in Algorithm 1,  6Rover degrades the obtained address patterns into a more generic form $P_g$ to represent a more fundamental method of address allocation. This step involves converting all non-zero parts of the address pattern into wildcards, accepting a larger scanning space in exchange for reducing the impact of pattern bias on the discovery of addresses in unseeded spaces. 

What's more, from line 8 to line 17, we iterate through the generic pattern list $P_g$ to perform preprocessing for constructing the Dependency Graph. First, from line 8 to line 12, we filter out generic patterns that do not appear in multiple prefixes or ASes. Given that Generic Patterns are already a broad representation of clustered states in address spaces, the likelihood of a pattern widely occurring in unseeded spaces is low when it does not appear in more than two ASes or prefixes. Second, from line 13 to line 17, we construct the mapping of the generic pattern to prefix $M$ based on the mapping of seed to prefix $H$, since the corresponding seed of each $pattern$ is recorded. This hash table $M$ is used as input to generate the dependency graph in line 18.

Finally, in line 18, 6Rover constructs the dependency graph between generic patterns. As 6Rover begins with probing patterns in smaller scanning spaces, upon discovering the applicability of a certain pattern in an unseeded space, it becomes necessary to identify other related and more complex patterns for further exploration. This relationship is constructed from two distinct perspectives. In the first case, when two generic patterns occur in the same prefix or AS and their wildcard counts differ by one, a dependency relationship is established, with the pattern having fewer wildcards pointing to the one with more. In the second case, if Generic Pattern $1$ is a subset of Generic Pattern $2$, we consider Generic Pattern $2$ to be dependent on Generic Pattern $1$. More specifically, Generic Pattern $2$ encompasses a larger scanning space that fully covers the scanning space of Generic Pattern $1$, as illustrated in the figure \ref{fig:dependency}. This indicates that when the applicability of Generic Pattern $1$ in a certain prefix is confirmed, directly related patterns with larger spaces also become worthwhile targets for exploration.
\input{Generic_Pattern_Extraction_pseudocode}

\subsection{Exploration \& Exploitation of Generic Pattern}
After the construction of Generic Patterns $P_g$ and Dependency Graph $D$, we have adopted an exploration-exploitation approach to optimize the selection of generic patterns and find the exact patterns applied in unseeded prefixes with fewer probe packets. Since the generalization of generic patterns through pattern degradation increases their applicability across different prefixes but also expands the probing space, we address this issue by employing the multi-armed bandit approach in reinforcement learning, achieving a balance between exploring unknown patterns and enhancing hit rates. 

\subsubsection{Main Algorithm Workflow}

Algorithm 3 and Algorithm 4 respectively illustrate the main process of Exploration \& Exploitation and the associated Utility Functions. 6Rover takes generic pattern list $G$, dependency graph $D$ and unseeded prefix set $N$ as input, finally outputs the active address set $S$ belonging to the unseeded prefixes. In Algorithm 3, 6Rover aims to progressively verify the usage of generic patterns in unseeded spaces according to the probing feedback. Consequently, for each unseeded prefix, 6Rover initiates multiple rounds of exploration \& exploitation, beginning with simple patterns and moving toward more complex ones. Once a relatively simple pattern is confirmed as applicable in the current space, the detection focus is expanded to more complex but related patterns based on the dependency graph of generic patterns. 

Firstly, from line 1 to line 4 in Algorithm 3, 6Rover initializes probed address hash map $H_a$ and probed pattern hash map $H_p$ to avoid the redundant probing of the same patterns or addresses. After that, 6Rover also initializes $G_{init}$ as the initial round of patterns to verify, including simple generic patterns with fewer than 4 wildcards.

Secondly, from line 5 to line 10, we enumerate each unseeded prefix from set $N$, and place all patterns from $G_{init}$ into a pattern queue $Q$, to serve as targets for the first round of multi-armed bandit exploration.

Thirdly, from line 11 to line 23, We first enter a loop relating to pattern queue $Q$. Since Q stores the set of patterns to be explored in the next round, the loop exits when the pattern set is empty. In line 13, After retrieving the set of patterns to be explored from $Q$, we input $Q$ and $H_a$ into the multi-armed bandit model. At this point, for ease of understanding, the multi-armed bandit model discussed later can be considered as a black box. Once the multi-armed bandit completes multiple rounds of exploration and exploitation, it outputs a set of effective patterns and active addresses. Additionally, from line 15 to line 18, based on dependency graph $D$ discussed in Section 4.2, 6Rover obtains related but more complex patterns for the next round of exploration.

\input{MAB}

\subsubsection{Multi-armed bandit model}

The multi-armed bandit model serves as the fundamental working unit and is repeatedly invoked by the main algorithm workflow. 6Rover applies the Upper Confidence Bound model\cite{auer2002finite} and continues iterative probing until reaching the maximum number of iterations or meeting exit criteria. Firstly, the arms of the multi-armed bandit are defined as the input of generic patterns from line 12 in Algorithm 3, which contains similar wildcard counts. Secondly, the reward function representing the feedback of each probing outcome is presented in 
formula (1), where $R_t(p)$ is the reward of choice pattern $p$ in iteration $t$, $S_t(p)$ is the active address count, and $A_t(p)$ is the aliased coefficient defined in formula (2).
\begin{equation}
    R_t(p) = S_t(p) \times A_t(p) 
\end{equation}

\begin{equation}
    A_t(p) = 
\begin{cases} 
    1 & \text{if } S_t(p) \text{ not aliased}, \\
    -\alpha & \text{if } S_t(p) \text{ aliased} .
\end{cases}
\end{equation}

The aliased coefficient $A_t(p)$ is employed to mitigate alias effects on detection direction. If the active address $S_t(p)$ is identified as alias addresses by the de-aliasing methods in Section 4.3.3, the original reward is multiplied by a negative amplification factor, $\alpha$, which is empirically fixed at 1, thereby avoiding persistent probing of the alias area and misleadingly high hit rate results.

After that, the multi-armed bandit model makes detection choices of generic patterns based on the formula (3), where $Q_t(p)$ is the estimation of action value of $p$ in time $t$, $c$ is a constant to control the balance of exploration and exploitation, and $N_t(p)$ is the number of choices of pattern $p$. After each detection of a generic pattern, $Q_t(p)$ is updated based on the reward $R_t(p)$ using the method outlined in Formula (4).

\begin{equation}
 A_t = \underset{p}{\mathrm{argmax}} \left( Q_t(p) + c \sqrt{\frac{\ln(t)}{N_t(p)}} \right)
\end{equation}

\begin{equation}
    Q_{t+1}(p) = \frac{Q_t(p) \times N_t(p) + R_t(p)}{N_t(p) + 1}
\end{equation}

Specifically, Algorithm 4 illustrates the main process of the multi-armed bandit model and the reward update mechanism. From line 1 to line 16, the multi-armed bandit model takes generic pattern $G$ and probed address hash map $H_a$ as input, and outputs the effective pattern list $G_{effective}$ and de-aliased active address set $S$.

Firstly, line 3 initializes bandit parameters including the maximum number of iterations $i_n$, upper bound confident constant $c$, Exiting threshold $k$, probing budget size $b$, hit rate threshold $r$, estimation of action value of each arm $Q$, number of times each arm has been picked $N$, and active address set $S$. From line 4 to line 7, the pre-scan procedure determines the initial rewards for each generic pattern.

After that, from line 8 to line 13, the multi-armed bandit continuously makes decisions based on Formula (3), updating the estimated rewards for generic patterns and collecting active addresses until either the maximum iteration count is reached or early termination criteria are met in line 9. We established a relatively low threshold of $k$ for early termination. If the average hit rate per action $\frac{\overline{Q}}{b}$ falls below this threshold, we infer a low probability of obtaining a significant number of active addresses in continued probing of that unseeded prefix which may be inactive or privacy-focused.

After completing a round of probing, 6Rover calculates the de-aliased hit rates for each generic pattern from line 14 to line 16. If a pattern’s hit rate exceeds the threshold $r$, it is considered applicable to the unseeded prefix. Based on the Dependency graph of this pattern, we further identify related patterns with one more wildcard. After filtering out patterns that have already been explored, these new patterns are used as arms for the next round of address discovery.

From line 17 to line 32 in Algorithm 4, 6Rover randomly samples IPv6 targets with probing budget size ratio $b$ from the chosen generic pattern $best_arm$. Since the multi-armed bandit may sample the same generic pattern multiple times across different iterations, to avoid redundant probing of the same addresses, we have maintained a hash table $H_a$ to store the activity status of each address. From line 28 to line 32, we utilize Formula (1), (2) and (4) to calculate rewards and update corresponding parameters.

\begin{table}[t]
\centering
\caption{Main Notations}
\begin{tabular}{lllr}
\toprule
\textbf{Notation} & \textbf{Definition}\\
\midrule
$G$     & Generic Pattern List                \\
$H_a$   & Probed Address Hash Map               \\
$S$   & Active Address Set               \\
$i_n$     & Maximum Number of Iterations         \\
$c$   & Upper Bound Confident Constant              \\
$k$   & Exiting Threshold for Early Termination in Each Round             \\
$r$     & Hit Rate Threshold for Effective Generic Patterns                \\
$b$   & Probing Budget Size               \\
$Q$   & Estimation of Action Value of Each Arms\\
$N$     & Number of Times Each Arm Has Been Picked   \\
$R$   & Reward of Probing Defined in Formula (1)               \\
$\alpha$   & Amplification Factor for De-aliasing    \\
\bottomrule
\end{tabular}
\end{table}
\subsubsection{Overcoming Performance Bottlenecks}
Employing multi-armed bandit mechanisms to control network probing presents two key bottleneck challenges. The first bottleneck challenge is the issue of probing efficiency, where the model continuously conducts sequential probing with a limited budget. In the multi-armed bandit mechanism, the choice of the next probing pattern is dependent on the outcome of the current probe. This necessitates a sequential process of 'selecting the best pattern,' 'probing,' 'updating the reward distribution of the probe pattern,' and then 're-selecting the best pattern,' thereby it is difficult to achieve parallelization to efficiently discover active targets. 

\begin{figure}[ht] 
    \centering 
    \includegraphics[width=0.85\linewidth]{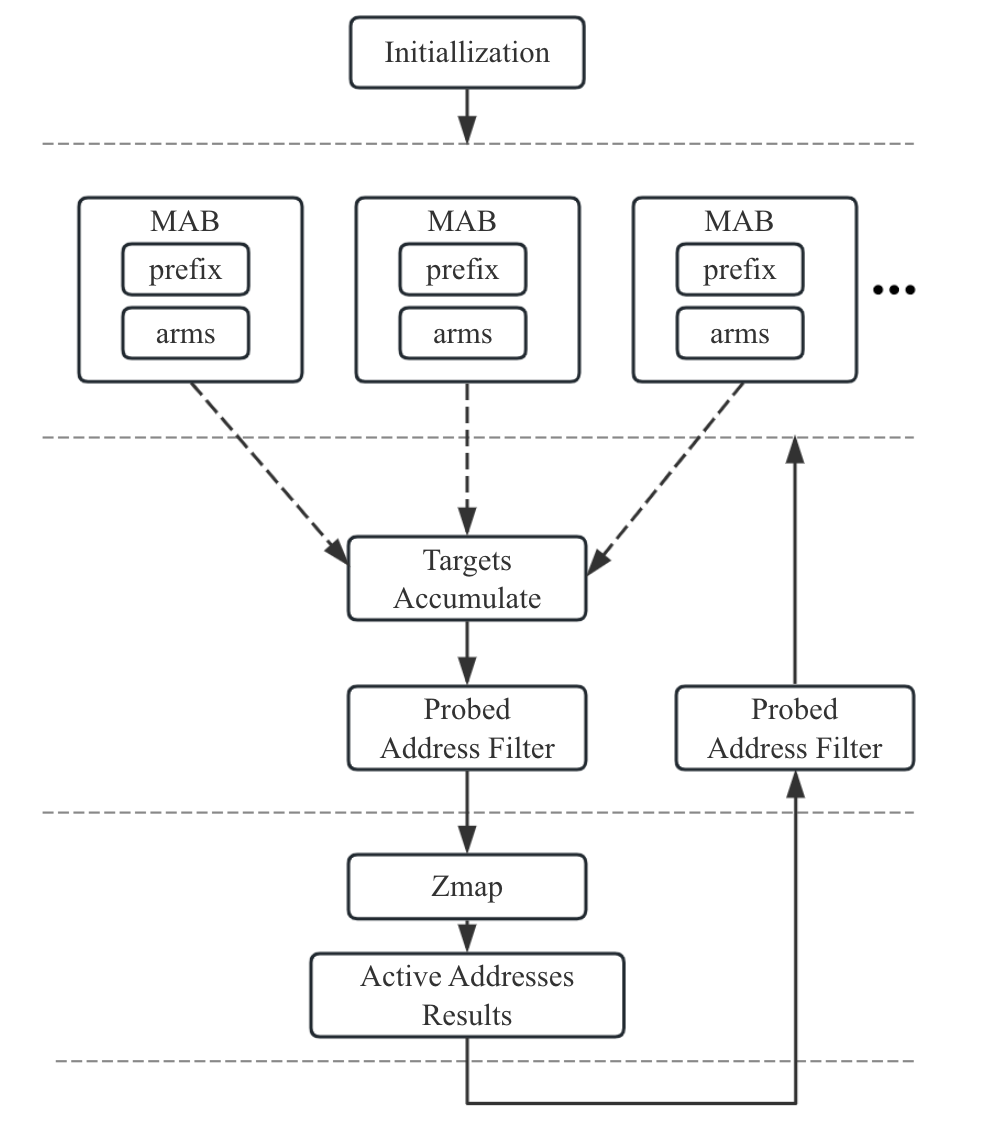} 
    \caption{The illustration of multi-armed bandit architecture. } 
    \label{fig:parallel_mab} 
\end{figure}

To overcome the efficiency bottleneck challenge, we introduce a multi-armed bandit architecture for parallel probing across multiple prefixes, as shown in Figure \ref{fig:parallel_mab}. This architecture achieves parallelization of probing by separating the decision and probing spaces of the multi-armed bandit model. Initially, multiple multi-armed bandit models are initialized, each with its independent prefix, pattern, and estimated reward distribution, operating independently in pattern selection and feedback reception. After each round of decision-making, the outcomes of these models are aggregated. An address filter maintains the association between probing addresses and model IDs, followed by unified probing using Zmap. Finally, active addresses identified in the probe are returned to their respective multi-armed bandit models through the address filter, enabling individual updates of model parameters. Hence, this architecture leverages Zmap's characteristic of higher efficiency in single-instance probing of a larger number of addresses, as opposed to several smaller probes, resulting in improved probing efficiency.

Additionally, the existence of aliasing addresses presents a significant challenge to the probing process in multi-armed bandit mechanisms. Due to the presence of alias IPs in IPv6 networks, which respond to probes to any IP within their alias prefix, there's a risk of the multi-armed bandit mechanism mistakenly allocating a large number of probes to these alias prefixes, leading to the illusion of superior probing performance. To solve this problem, we implemented a stringent alias detection mechanism. Initially, when initializing prefixes, we exclude those listed in the Gasser continually updated alias prefix dataset \cite{ipv6_hitlist} and also conduct a pre-scan of 10 randomly generated addresses for each prefix. Prefixes with active addresses detected during this pre-scan are marked as alias prefixes and are not used for further probing. Furthermore, when the multi-armed bandit suggests a pattern for probing, we generate and pre-scan five addresses fitting the pattern. If all these addresses are active, it indicates the triggering of aliasing addresses, and the model receives a significantly negative reward to prevent repeat probing of that pattern.

\section{Evaluation}
\subsection{Experiment Setup}
\subsubsection{Dataset Description}
To construct a comprehensive set of generic patterns, 6Rover integrates hitlists from three data sources, merging a dataset of 100 million historical active addresses. These sources consist of the Gasser historical hitlist from October 2021 to November 2023, the publicly available HMap6, and the Addrminer hitlist. The basic information of these datasets is shown in Table 2. 

\begin{table}[h]
\centering
\caption{Hitlist Dataset Comparison}
\begin{tabular}{lllr}
\toprule
\textbf{Dataset} & \textbf{\#Seeds} & \textbf{\#prefixes} & \textbf{\#ASes} \\
\midrule
Gasser Hitlist     & 42.3M    & 66.9k   & 20.3k            \\
Addrminer   & 22.4M    & 54.7k & 17.7k             \\
HMap6   & 126.4M    & 55.3k & 18.7k             \\
\midrule
Total  & 160.5M & 78.6k & 21.8k \\
\bottomrule
\end{tabular}
\end{table}

Additionally, Figure 2 presents the distribution of address counts across various prefixes for the three data sources. Due to the considerable differences in the number of prefixes and addresses among these sources, for better visualization and comparison of the three datasets, we normalized the address counts on the vertical axis and implemented linear interpolation for the number of prefixes on the horizontal axis.
\begin{figure*}[ht]
    \centering
    \includegraphics[width=0.4\textwidth]{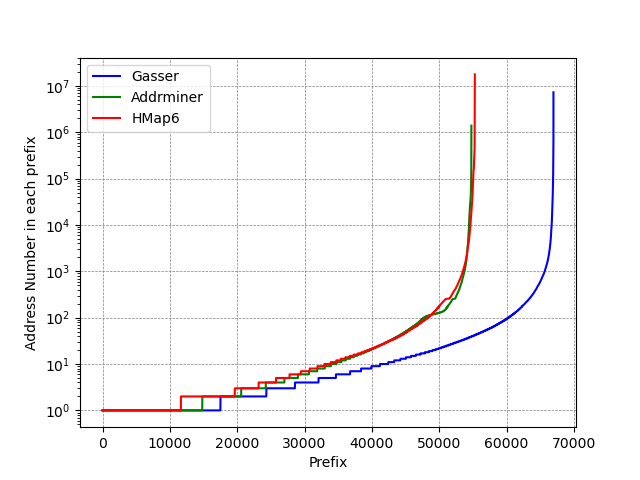}
    \includegraphics[width=0.405\textwidth]{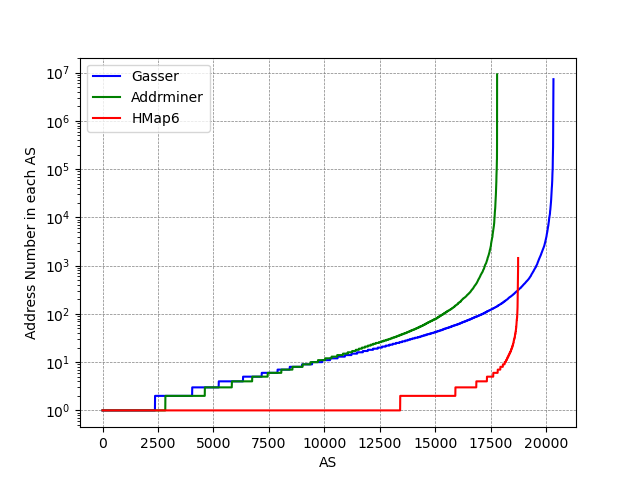}
    \caption{distribution of seed addresses}
    \label{fig:yourlabel}
\end{figure*}

Figure 2 clearly shows that despite the datasets comprising a more comprehensive collection of historical active addresses, the distribution of active addresses across various prefixes is highly uneven in each dataset. The seed address distribution in Figure 2 indicates that, even in seeded areas, approximately 25\% of prefixes contain only one address, and about 50\% of prefixes have fewer than ten addresses. Due to this scarcity of addresses, algorithms for address generation within seeded spaces are likely to be less effective. 

Consequently, in our subsequent experiments, we constructed three types of datasets representing unseeded spaces, including prefixes not appearing in the hitlist, prefixes in the hitlist with fewer than 20 seeds, and prefixes present in the hitlist but with all addresses being inactive. We randomly sampled 10000 prefixes from each type and compared the hit rates of addresses generated by 6Rover and other methods within these prefixes, and we rigorously followed the ethical guidelines recommended by Partridge et al. \cite{ethic1} and Dittrich et al.\cite{ethic2} to prevent any interference with the regular operation of devices in the IPv6 network.

\subsubsection{System Parameters}
All experiments and method implementations were conducted on an Alibaba Cloud Linux server equipped with 24GB of memory, an Intel Xeon Platinum 8163 CPU, and a bandwidth of 100Mbps. Within the multi-armed bandit model, we set the maximum number of iterations to 100 times the number of arms, the upper bound confidence constant at 50, the exiting threshold at 2.5\%, the probing budget size ratio at 0.1, and the hit rate threshold at 5\%. Comparative experiments for parameter optimization of these settings are detailed in Section 5.3.

\subsection{Real-world Test Results}
\subsubsection{Hit Rate Performance}
In this section, we compare 6Rover with the state-of-the-art unseeded prefix address discovery methods, Addrminer\cite{addrminer} and HMap6\cite{hmap6}, under the same experimental conditions. Apart from the pattern library construction process of Addrminer and 6Rover, which utilized addresses from seeded spaces, we ensured that no active address information was leaked during the probing process in unseeded spaces. From the three types of unseeded spaces defined in section 5.1.1 including sparse hitlist prefixes, inactive hitlist prefixes, and unlisted prefixes, we randomly selected 10000 non-alias prefixes as the target space for probing.
\begin{figure}[ht]
    \centering
    \includegraphics[width=0.3\textwidth]{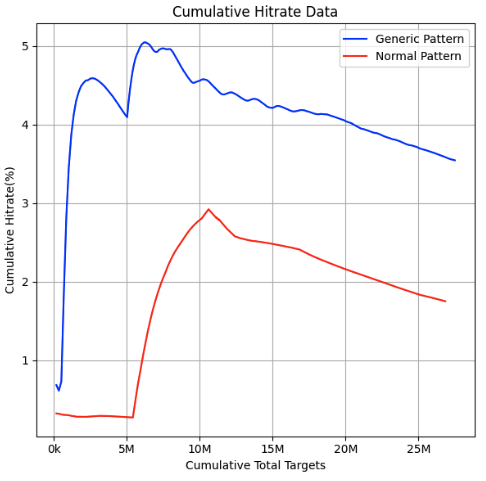}
    \caption{Performance Comparison between Generic Patterns and Normal Patterns}
    \label{fig:pattern}
\end{figure}

\begin{figure}[ht]
    \centering
    \includegraphics[width=0.4\textwidth]{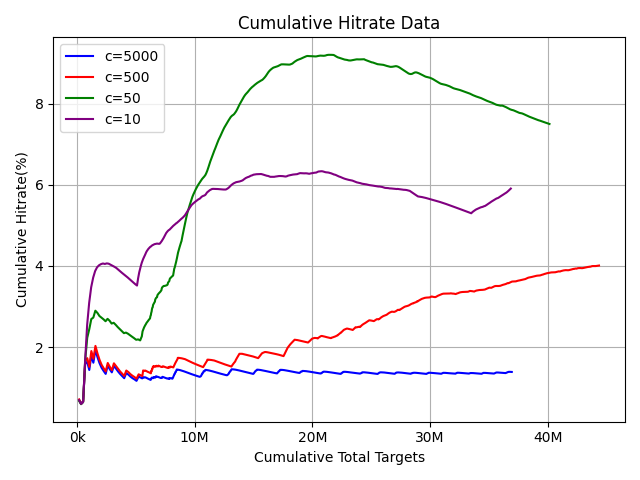}
    \caption{Performance Comparison between Generic Patterns and Normal Patterns}
    \label{fig:ctest1}
\end{figure}

In Figure 7, the hit rate results of 6Rover and its baseline counterpart are presented for three types of unseeded prefix datasets. Given that 6Rover is based on a dynamic scanning model utilizing reinforcement learning, to better illustrate the variations in each round of probing, we present the hit rate curves in chronological order. Meanwhile, as Addrminer and HMap6 do not specify a probing sequence, their results are exhibited in a sorted format. In each of the three scenarios, 6Rover exhibits a more effective hit rate performance, reaching about a 7\% cumulative hit rate at a budget size of 40M, thus exceeding the performance of state-of-the-art methods. 

Moreover, it's noticeable that Addrminer and HMap6 show extremely high hit rates in a small subset of prefixes. For Addrminer, this could be attributed to the model successfully identifying prefixes with significant correlation through WHOIS data comparisons, hence migrating patterns from seeded to unseeded spaces and achieving higher hit rates. However, in cases where the WHOIS information of prefixes is completely unassociated, the algorithms lean towards a random pattern-matching approach from the standard library, leading to reduced hit rates. Similarly, HMap6's heuristic methodology can accurately match addresses in numerous prefixes following established empirical rules. Yet, this approach falls short in detecting addresses in patterns that diverge from these rules, thus diminishing the overall hit rate.

\begin{figure*}[ht]
    \centering
    \includegraphics[width=0.8\textwidth]{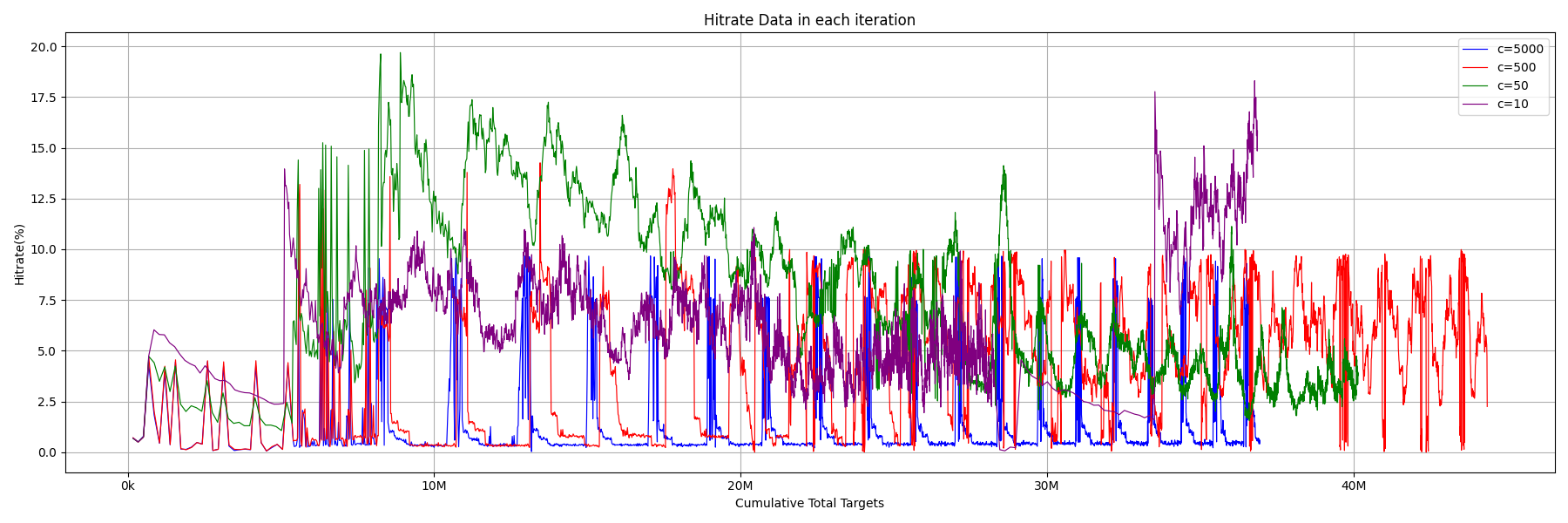}
    \caption{Performance Comparison between Generic Patterns and Normal Patterns}
    \label{fig:ctest2}
\end{figure*}

\begin{figure*}[ht]
    \centering
    \subfloat[Sparse Hitlist Prefixes]{%
        \includegraphics[width=0.26\textwidth]{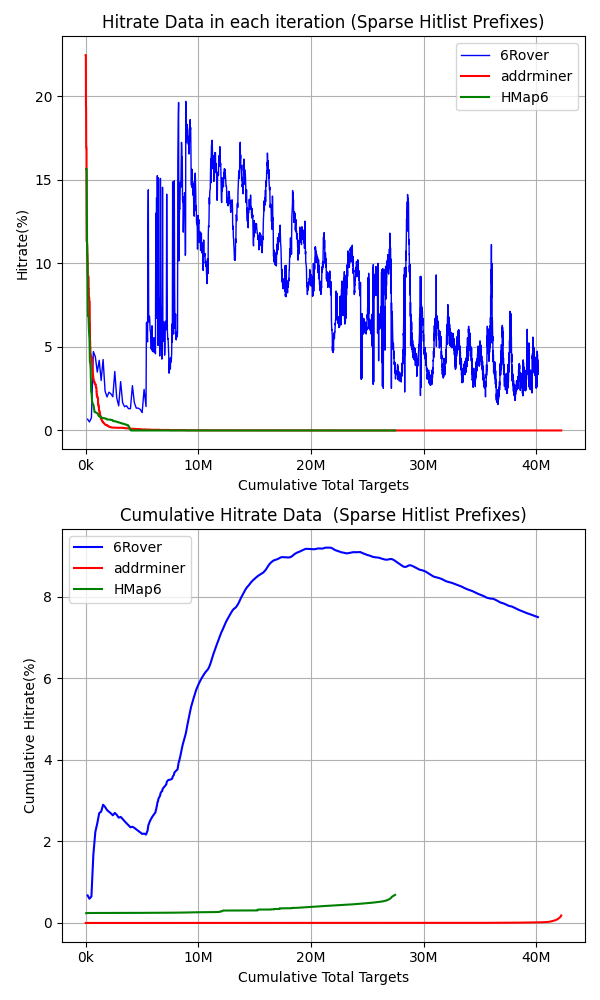}
        \label{fig:active_baseline}
    }
    \hfill 
    \subfloat[Inactive Hitlist Prefixes]{%
        \includegraphics[width=0.26\textwidth]{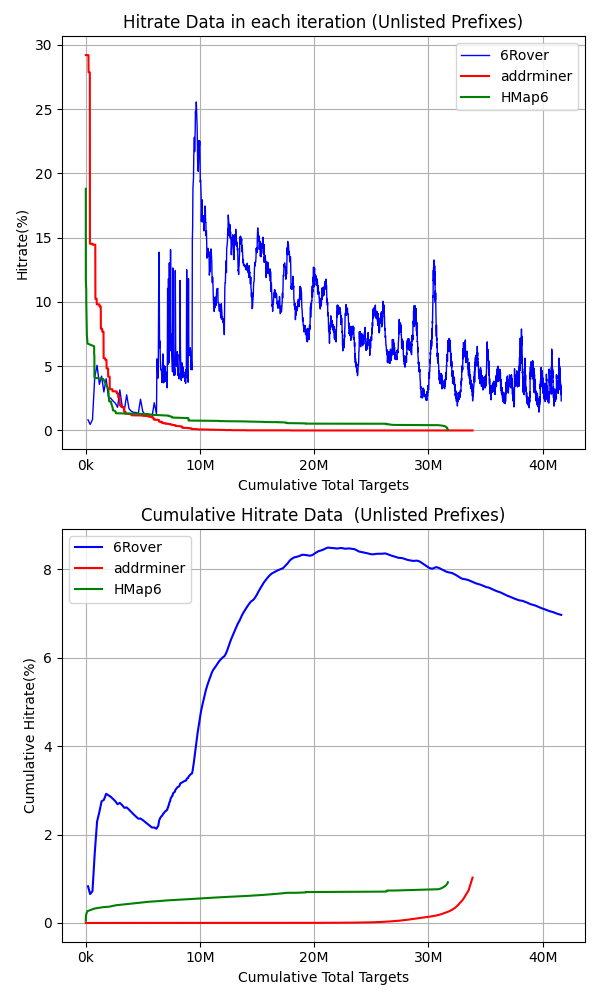}
        \label{fig:inactive_baseline}
    }
    \hfill 
    \subfloat[Unlisted Prefixes]{%
        \includegraphics[width=0.26\textwidth]{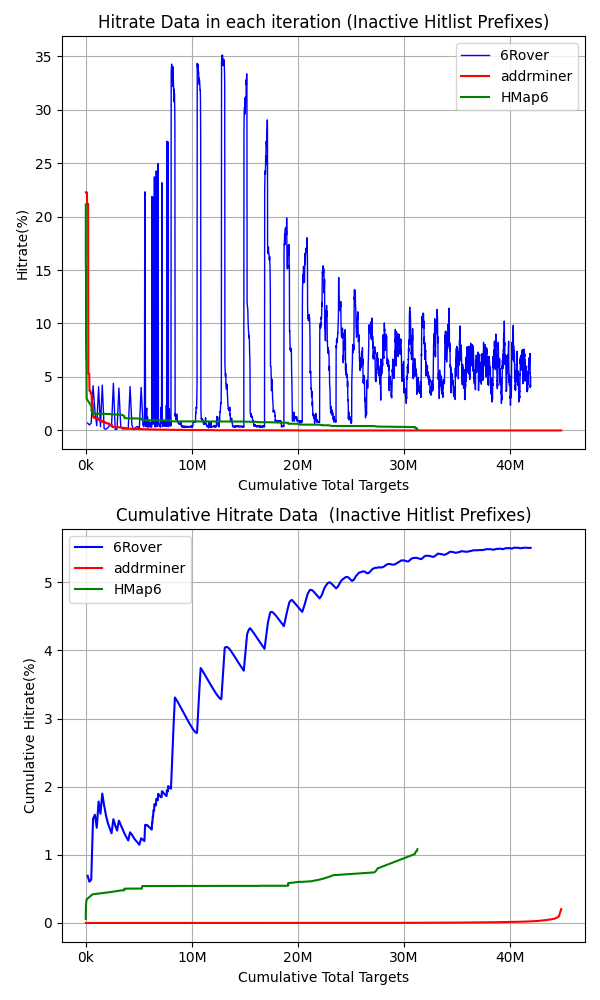}
        \label{fig:seedless_baseline}
    }
    \caption{Hit rate performances}
    \label{fig:three_images}
\end{figure*}

In each dataset, we observed a trend in hit rate that initially increases, then decreases, which can be attributed to 6Rover's exploration-exploitation model. It gradually constructs the reward distribution for each candidate pattern through step-by-step exploration, before transitioning to the exploitation phase. Consequently, during the early exploration stage, the hit rate is lower, but as the reward distribution becomes clearer, the model shifts more towards exploitation, leading to a gradual increase in the hit rate. Additionally, Furthermore, the gradual decrease in hit rate can be attributed to 6Rover's strategy of starting with simpler patterns and progressively moving to more complex ones. With the increased complexity of patterns, the probing space for each becomes larger, resulting in lower hit rates. However, it still sustains a consistent hit rate of about 3\% - 5\% in each dataset, which is significantly higher than that of Addrminer and HMap6 at the same budget size.

When comparing 6Rover's performance across the three datasets, we observe that its performance exhibits similar characteristics in both the Spare Hitlist Prefixes and Inactive Hitlist Prefixes datasets, with a higher hit rate than in the unlisted prefixes dataset. This suggests that prefixes featured in the hitlist, even with few or no currently active addresses, can still be probed effectively by 6Rover to uncover more potential active addresses, owing to their confirmed previous usage. Moreover, within the Unlisted Prefixes dataset, 6Rover displays a distinct pattern in its probing outcomes per round, unlike the other two datasets. Specifically, before reaching 30M, there are rounds with both lower and higher hit rates, resulting in a large variability in the overall hit rate. This is attributed to the presence of certain prefixes in the unlisted prefixes dataset where active addresses cannot be detected, leading to lower hit rates in some rounds. Such prefixes may either be inactive or have strong privacy features, such as client addresses using dynamic IPs configured by SLAAC \cite{SLAAC}.

\subsubsection{Parameter Optimization and Evaluation}

Figure \ref{fig:pattern} illustrates the effectiveness of Generic Patterns through a comparison with Normal Patterns. In this experiment, both pattern types were integrated into an identical multi-armed bandit probing architecture with the same parameters, and deployed on servers with the same configurations to probe the same collection of unseeded prefixes. Results indicate that the use of Generic Patterns in the multi-armed bandit model leads to enhanced probing efficiency, achieving an improvement of about 2\% compared to Normal Patterns. This observation can be interpreted as normal patterns distinctly mapping out hotspots in seeded prefixes, but such precise hotspots may not exist in unseeded spaces. In contrast, generic patterns afford a wider scope for exploration, which, aided by the multi-armed bandit framework, leads to the discovery of more active addresses in unseeded prefixes.

Additionally, in the UBC-based multi-armed bandit model, a critical parameter influencing the probing effectiveness is the constant $c$, which governs the balance between exploration and exploitation. To determine the optimal value of $c$ for peak performance, we conducted comparative experiments on various values of $c$. Figures \ref{fig:ctest1} and \ref{fig:ctest2} illustrate the performance of parameter $c$ at different settings. As seen in Figure \ref{fig:ctest1}, the model achieves its highest hit rate with $c$ valued at 50, striking an optimal balance between exploration and exploitation. When $c$ is significantly higher, for instance at 5000 and 1000, the early hit rate patterns appear comparable. This is because an overly large $c$ value excessively emphasizes exploration over exploitation, causing the model to employ a sequential pattern selection approach for total exploration.

\section{Conclusion}
This work introduces a method, 6Rover, which efficiently discoverys active IPv6 addresses without relying on any seed address information within the current space. By generating patterns with higher generality that reflect true network administrator strategies, 6Rover efficiently discovers active addresses without prior seed knowledge. Employing a multi-armed bandit model, it optimally allocates probing resources, addressing the exploration-exploitation dilemma in unseeded space pattern discovery. Empirical results demonstrate 6Rover's superior performance, achieving a 5\% - 8\% hit rate in seedless environments with a 100 million budget, an approximately 200\% improvement over current state-of-the-art methods. This significant advancement in IPv6 network surveying holds promise for enhanced network topology measurements and security analysis.


%




\section*{Acknowledgments}
This work was supported in part by a grant from NSFC Grant no. 62101159, NSF of Shandong Grant no. ZR2021MF055, and also the Research Grants Council of Hong Kong under the Areas of Excellence scheme grant AoE/E-601/22-R. 

\bibliographystyle{IEEEtran}
\bibliography{references} 


\newpage

\vspace{11pt}

\vspace{-33pt}
\begin{IEEEbiography}[{\includegraphics[width=1in,height=1.25in,clip,keepaspectratio]{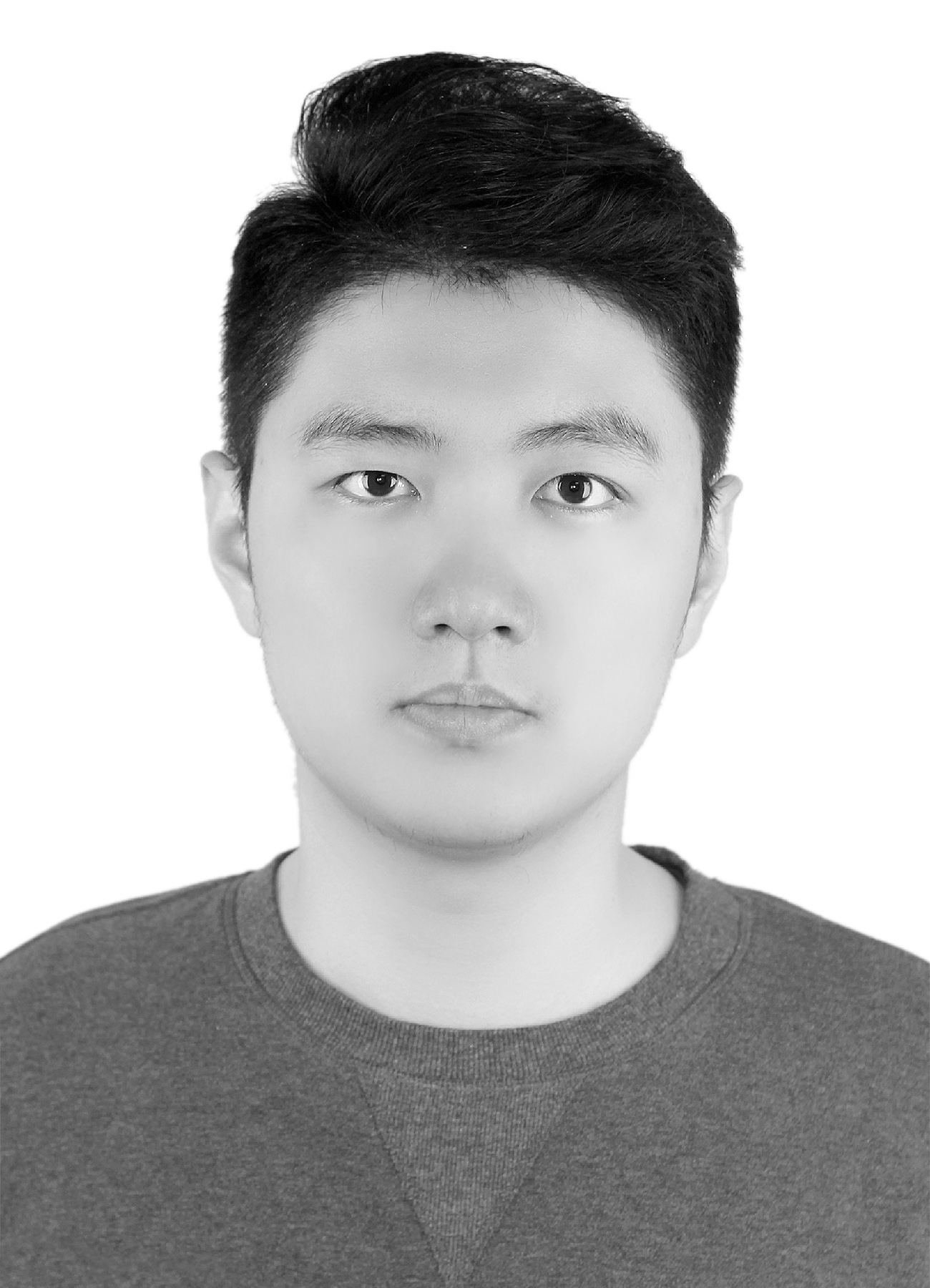}}]{Zhichao Zhang}
received his B.S. degree from the University of Electronic Science and Technology of China, Chengdu in 2019, and his M.S. degree from Columbia University, New York in 2021. He is currently pursuing a Ph.D. degree with the Harbin Institute of Technology, Harbin, China. His research interest includes IPv6 network measurements and network security.
\end{IEEEbiography}
\begin{IEEEbiography}[{\includegraphics[width=1in,height=1.25in,clip,keepaspectratio]{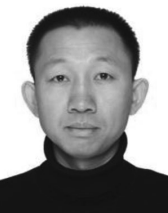}}]{Zhaoxin Zhang}
received the PhD degree from the Harbin Institute of Technology, in 2007. Since
2007 he has been employed with the School of Computer Science and Technology, Harbin Institute of Technology, Weihai. Now he is a professor there. And since 2007 he has been in charge of
the Network and Information Security Technology Research Center of Harbin Institute of Technology at Weihai. His research interests include focuses on network and information security, network simulation, and domain name system security.
\end{IEEEbiography}
\begin{IEEEbiography}[{\includegraphics[width=1in,height=1.25in,clip,keepaspectratio]{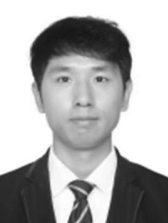}}]{Yanan Cheng}
received the Ph.D. degree from the Harbin Institute of Technology, Harbin, China, in 2023. He is currently a Lecturer with the Faculty of Computing, Harbin Institute of Technology. His research interests include network and information security and domain name system security.
\end{IEEEbiography}
\begin{IEEEbiography}[{\includegraphics[width=1in,height=1.25in,clip,keepaspectratio]{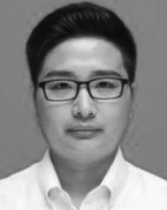}}]{Ning Li}
(Member, IEEE) received the BS and MS degrees from Zhengzhou University, China, in 2010 and 2014, respectively, and the PhD degree in telematic engineering from the Universidad Politecnica de Madrid (UPM), Spain, in 2018. He is currently an assistant professor with the School of Computer Science and Technology, Harbin Institute of Technology, China. He has authored many journal papers (SCI indexed). His main research interests include the areas of mobile ad hoc networks, network protocols, topology control, network optimization, and distributed computing.
\end{IEEEbiography}

\vfill

\end{document}

%% file: Generic_Pattern_Extraction_pseudocode.tex
\begin{algorithm}[H]
\caption{Generic Pattern Construction}
\begin{algorithmic}[1]
\small
\Require  The seed set $C$ , the seed to prefix map $H$ 
\Ensure  The generic pattern list $G$, the dependency graph $D$ 
\State  $C_t, H_t \gets SeedTruncate(C, H)$ \Comment{Truncate the address to remove its global routing prefix.}
\State $P \gets PatternMining(C_t)$ \Comment{Employ existing Pattern Mining Algorithm such as 6Graph, 6Forest and DET, to construct pattern set from seed addresses }
\State \textbf{Initialize} generic pattern list $P_g$,  generic pattern to prefix map $M$
\For{$pattern \in P$}
    \State $pattern_g \gets PatternDegrade(pattern)$\
    \State $P_g \gets P_g \cup \{pattern_g\}$
\EndFor
\For{$pattern \in P_g$}
    \If {not $IsSharedPattern(pattern)$} \Comment{Maintain only the generic patterns that are shared across multiple ASes.}
    \State $P_g.remove(pattern)$
    \State \textbf{continue}
    \EndIf
    \For {$seed \in pattern.seeds$}  \Comment{During the construction of patterns, the related seeds are recorded in this field}
        \State $prefix \gets H_t[seed]$
        \State $M[pattern_g] \gets M[pattern_g] \cup \{prefix\}$
    \EndFor
\EndFor
\State $D \gets ConstructDependency(P_g, M)$
\State \Return $G, D$
\Statex
\end{algorithmic}
\end{algorithm}

\begin{algorithm}[H]
\small
\caption{Utility Functions for Generic Pattern Construction}
\begin{algorithmic}[1]
\Function{\textbf{S}eed\textbf{T}runcate}{$C$, $H$}
    \State \textbf{initialize} truncated seed set $C_t$, truncated seed to prefix map $H_t$
    \For{$seed \in C$}
        \State $seed_t \gets seed[48:128]$
        \State \( C_t \gets C_t \cup \{seed_t\} \)
        \State $prefix \gets H[seed]$
        \State $H_t[seed_t] \gets H_t[seed_t] \cup \{prefix\}$
    \EndFor
    \State \Return $C_t, H_t$
\EndFunction
\Statex
\Function{\textbf{P}attern\textbf{D}egrade}{$pattern$}
    \For{$i \gets 0 \textrm{ to } len(pattern)-1$}
        \If{$pattern[i] \neq "0" $ }
            \State $pattern[i] \gets "*"$
        \EndIf
    \EndFor
    \State \Return $pattern$
\EndFunction
\Statex

\Function{\textbf{C}onstruct\textbf{D}ependency}{$P_g, M$}
    \State \textbf{initialize} dependency graph D
    \For{$i \gets 0 \textrm{ to } len(P_g)-1$}
        \State $D.addNode(P_g[i])$
    \EndFor
    \For{$node1 \in P_g$}
        \For{$node2 \in P_g$}
            \If{$node1 == node2 \textbf{ or } node1.count("*") > ndoe2.count("*")$}
                \State \textbf{continue}
            \EndIf
            \If{$M[node1] == M[node2]$} 
                \State $D.addEdge(node1, node2)$ 
            \EndIf
            \If{$IsSubPattern(node1, node2)$}
                \State $D.addEdge(node1, node2)$ 
            \EndIf
                        
        \EndFor
    \EndFor
    \State \Return $pattern$
\EndFunction
\Statex

\Function{\textbf{I}s\textbf{S}ub\textbf{P}attern}{$node1, node2$}
    \For{$i \gets 0 \textrm{ to } len(node1)-1$}
        \If{$node1[i] == "*" \textbf{ and } node2[i] == "0" $ }
            \State \Return \textbf{False}
        \EndIf
    \EndFor
    \State \Return \textbf{True}
\EndFunction
\Statex
\end{algorithmic}
\end{algorithm}

%% file: MAB.tex
\begin{algorithm}
\small
\caption{Exploration \& Exploitation of Generic Patterns}
\begin{algorithmic}[1]
\Require  The generic pattern list $G$, the dependency graph $D$, the unseeded prefix set $N$
\Ensure  Active Address Set $S$
\State \textbf{Initialize} Probed Address Hash Map $H_a$, Probed Pattern Hash Map $H_p$
\State $G_{init} \gets \{e \text{ for } e \in G \text{ if } wildcard.count(e) == 3\}$
\State $G_{small} \gets \{e \text{ for } e \in G \text{ if } wildcard.count(e) \leq 2 \}$
\State $G_{init} \gets G_{init} \cup {\textbf{merge}(G_{small})}$ \Comment{Generate the initial patterns}

\For{$\text{each } prefix \in N$}
    \State \textbf{Initialize} Pattern Queue $Q$
    \For{$\text{each } init\_pattern \in G_{init}$}
        \State $H_p.add(init\_pattern)$
        \State $Q.enqueue(init\_pattern)$
    \EndFor
    \While{$Q$ is not empty}
    \State $G_{temp} \gets Q.getall()$ \Comment{Pop all patterns from queue}
    \State $G_{effective}, active\_addrs \gets MAB(G_{temp}, H_a)$ \Comment{MAB is the abbreviation of MultiArmedBandit Function}
    \State $S \gets S \cup active\_addrs$
    \For{$pattern \in G_{effective}$}
        \If{$pattern \notin H_p $}
            \State $pattern_{next} \gets D[pattern]$ 
            \State $Q.enqueue(pattern_{next})$
        \EndIf
\EndFor
\EndWhile
\EndFor
\State \textbf{Return} S
\Statex
\end{algorithmic}
\end{algorithm}

\begin{algorithm}
\small
\caption{Utility Functions for Exploration \& Exploitation of Generic Patterns}
\begin{algorithmic}[1]
\Function{\textbf{M}ULTI\textbf{A}RMED\textbf{B}ANDIT}{$G$, $H_a$}
    \State \textbf{initialize} input parameters: pattern list $G$, Probed Address Hash Map $H_a$
    \State \textbf{initialize} bandit parameters:  $i_n$ ,  $c$,  $k$,  $r$, $Q$, $N$, $S$, $b$
    \For{each $p \in G$}
        \State $Q, N, H_a, S_{temp} \gets RewardUpdate(p, Q, N, H_a, b)$  \Comment{Initialize $Q$ and $N$ by pre-scan each patterns once}
        \State $S \gets S \cup S_{temp}$
    \EndFor
    \State $t \gets 0$
    \While{$t < i_n \textbf{ and } \frac{\overline{Q}}{b}  > k $} 
       \State $best\_arm \gets \underset{p}{\mathrm{argmax}} \left( Q_t(p) + c \sqrt{\frac{\ln(t)}{N_t(p)}} \right)$

       \State $Q, N, H_a ,  S_{temp}\gets RU(best\_arm, Q, N, H_a, b)$
       \Comment{RU is the abbreviation of RewardUpdate Function}
       \State $S \gets S \cup S_{temp}$
    \EndWhile
    \State $G_{effective} \gets \{g \textbf{ for } g \in Q \textbf{ if } \frac{Q[g]}{b} \geq r\}$
    \State \Return $G_{effective}, S$
\EndFunction
\Statex
\Function{\textbf{R}EWARD\textbf{U}PDATE}{$best\_arm, Q, N, H_a, b$}
    \State \textbf{initialize} active address set $S_{temp}$
   \State $targets \gets randomSample(best\_arm)$ \Comment{Random sample addresses from a given pattern}
   \For{each $target \in targets$ }
    \If{$target \in H_a$}
        \State $targets.del(target)$, $S_{temp}.add(target)$
        \State \textbf{continue}
    \EndIf
        \State $M_a[target] \gets True$
   \EndFor
    \State $active\_addr \gets Scan(targets,  b)$ \Comment{Probing de-aliased active addresses in real network}
    \State $Q[best\_arm] \gets \text{Update Rewards by Formula(1)(2)(4)}$
    
    \State $N[best\_arm] \gets N[best\_arm] + 1$
    \State $S_{temp} \gets S_{temp} \cup active\_addr$
    \State \Return $Q, N, H_a, S_{temp}$
\EndFunction
\Statex

\end{algorithmic}
\end{algorithm}

%% file: bare_jrnl_new_sample4.bbl
\begin{thebibliography}{10}
\providecommand{\url}[1]{#1}
\csname url@samestyle\endcsname
\providecommand{\newblock}{\relax}
\providecommand{\bibinfo}[2]{#2}
\providecommand{\BIBentrySTDinterwordspacing}{\spaceskip=0pt\relax}
\providecommand{\BIBentryALTinterwordstretchfactor}{4}
\providecommand{\BIBentryALTinterwordspacing}{\spaceskip=\fontdimen2\font plus
\BIBentryALTinterwordstretchfactor\fontdimen3\font minus \fontdimen4\font\relax}
\providecommand{\BIBforeignlanguage}[2]{{%
\expandafter\ifx\csname l@#1\endcsname\relax
\typeout{** WARNING: IEEEtran.bst: No hyphenation pattern has been}%
\typeout{** loaded for the language `#1'. Using the pattern for}%
\typeout{** the default language instead.}%
\else
\language=\csname l@#1\endcsname
\fi
#2}}
\providecommand{\BIBdecl}{\relax}
\BIBdecl

\bibitem{google_v6_stats}
\BIBentryALTinterwordspacing
Google, ``Ipv6 adoption statistics,'' 2023. [Online]. Available: \url{https://www.google.com/intl/en/ipv6/statistics.html}
\BIBentrySTDinterwordspacing

\bibitem{10.1145/3278532.3278559}
\BIBentryALTinterwordspacing
R.~Beverly, R.~Durairajan, D.~Plonka, and J.~P. Rohrer, ``In the ip of the beholder: Strategies for active ipv6 topology discovery,'' in \emph{Proceedings of the Internet Measurement Conference 2018}, ser. IMC '18.\hskip 1em plus 0.5em minus 0.4em\relax New York, NY, USA: Association for Computing Machinery, 2018, p. 308–321. [Online]. Available: \url{https://doi.org/10.1145/3278532.3278559}
\BIBentrySTDinterwordspacing

\bibitem{JIA2019106947}
\BIBentryALTinterwordspacing
S.~Jia, M.~Luckie, B.~Huffaker, A.~Elmokashfi, E.~Aben, K.~Claffy, and A.~Dhamdhere, ``Tracking the deployment of ipv6: Topology, routing and performance,'' \emph{Computer Networks}, vol. 165, p. 106947, 2019. [Online]. Available: \url{https://www.sciencedirect.com/science/article/pii/S1389128618304092}
\BIBentrySTDinterwordspacing

\bibitem{6search}
\BIBentryALTinterwordspacing
N.~Liu, C.~Jia, B.~Hou, C.~Hou, Y.~Chen, and Z.~Cai, ``6search: {A} reinforcement learning-based traceroute approach for efficient ipv6 topology discovery,'' \emph{Comput. Networks}, vol. 235, p. 109987, 2023. [Online]. Available: \url{https://doi.org/10.1016/j.comnet.2023.109987}
\BIBentrySTDinterwordspacing

\bibitem{gws-geo}
\BIBentryALTinterwordspacing
Z.~Ma, S.~Zhang, X.~Hu, N.~Li, Q.~Zhou, F.~Liu, H.~Wang, G.~Hu, and Q.~Dong, ``Gws-geo: {A} graph neural network based model for street-level ipv6 geolocation,'' \emph{J. Inf. Secur. Appl.}, vol.~75, p. 103511, 2023. [Online]. Available: \url{https://doi.org/10.1016/j.jisa.2023.103511}
\BIBentrySTDinterwordspacing

\bibitem{hgl-geo}
Z.~Ma, X.~Hu, N.~Li, H.~Feng, S.~Zhang, T.~Li, F.~Liu, Q.~Zhou, Z.~Tian, H.~Wang, and G.~Hu, ``Hgl-geo: Finer-grained ipv6 geolocation algorithm based on hypergraph learning,'' \emph{Information Processing \& Management}, vol.~60, no.~6, p. 103518, 2023.

\bibitem{ipvseeyou}
E.~C. Rye and R.~Beverly, ``Ipvseeyou: Exploiting leaked identifiers in ipv6 for street-level geolocation,'' in \emph{2023 IEEE Symposium on Security and Privacy (SP)}, 2023, pp. 3129--3145.

\bibitem{durumeric2013zmap}
Z.~Durumeric, E.~Wustrow, and J.~A. Halderman, ``$\{$ZMap$\}$: fast internet-wide scanning and its security applications,'' in \emph{22nd USENIX Security Symposium (USENIX Security 13)}, 2013, pp. 605--620.

\bibitem{izhikevich2021lzr}
L.~Izhikevich, R.~Teixeira, and Z.~Durumeric, ``$\{$LZR$\}$: Identifying unexpected internet services,'' in \emph{30th USENIX Security Symposium (USENIX Security 21)}, 2021, pp. 3111--3128.

\bibitem{li2021fast}
X.~Li, B.~Liu, X.~Zheng, H.~Duan, Q.~Li, and Y.~Huang, ``Fast ipv6 network periphery discovery and security implications,'' in \emph{2021 51st Annual IEEE/IFIP International Conference on Dependable Systems and Networks (DSN)}.\hskip 1em plus 0.5em minus 0.4em\relax IEEE, 2021, pp. 88--100.

\bibitem{li2020towards}
F.~Li and D.~Freeman, ``Towards a user-level understanding of ipv6 behavior,'' in \emph{Proceedings of the ACM Internet Measurement Conference}, 2020, pp. 428--442.

\bibitem{10.1145/3544912.3544915}
\BIBentryALTinterwordspacing
S.~J. Saidi, O.~Gasser, and G.~Smaragdakis, ``One bad apple can spoil your ipv6 privacy,'' \emph{SIGCOMM Comput. Commun. Rev.}, vol.~52, no.~2, p. 10–19, jun 2022. [Online]. Available: \url{https://doi.org/10.1145/3544912.3544915}
\BIBentrySTDinterwordspacing

\bibitem{rye2021follow}
E.~Rye, R.~Beverly, and K.~C. Claffy, ``Follow the scent: Defeating ipv6 prefix rotation privacy,'' in \emph{Proceedings of the 21st ACM Internet Measurement Conference}, 2021, pp. 739--752.

\bibitem{6gen}
\BIBentryALTinterwordspacing
A.~Murdock, F.~Li, P.~Bramsen, Z.~Durumeric, and V.~Paxson, ``Target generation for internet-wide ipv6 scanning,'' in \emph{Proceedings of the 2017 Internet Measurement Conference}, ser. IMC '17.\hskip 1em plus 0.5em minus 0.4em\relax New York, NY, USA: Association for Computing Machinery, 2017, p. 242–253. [Online]. Available: \url{https://doi.org/10.1145/3131365.3131405}
\BIBentrySTDinterwordspacing

\bibitem{barnes2012mapping}
R.~Barnes, R.~Altmann, and D.~Kerr, ``Mapping the great void: Smarter scanning for ipv6,'' \emph{Proc. CAIDA AIMS-4}, 2012.

\bibitem{pam2017-something-from-nothing-there}
T.~Fiebig, K.~Borgolte, S.~Hao, C.~Kruegel, and G.~Vigna, ``{Something From Nothing (There): Collecting Global IPv6 Datasets From DNS},'' in \emph{Proceedings of the 18th Passive and Active Measurement (PAM)}, ser. Lecture Notes in Computer Science (LNCS), M.~A. Kâafar, S.~Uhlig, and J.~Amann, Eds., vol. 10176.\hskip 1em plus 0.5em minus 0.4em\relax Springer International Publishing, pp. 30--43.

\bibitem{DBLP:conf/sp/BorgolteHFV18}
\BIBentryALTinterwordspacing
K.~Borgolte, S.~Hao, T.~Fiebig, and G.~Vigna, ``Enumerating active ipv6 hosts for large-scale security scans via dnssec-signed reverse zones,'' in \emph{2018 {IEEE} Symposium on Security and Privacy, {SP} 2018, Proceedings, 21-23 May 2018, San Francisco, California, {USA}}.\hskip 1em plus 0.5em minus 0.4em\relax {IEEE} Computer Society, 2018, pp. 770--784. [Online]. Available: \url{https://doi.org/10.1109/SP.2018.00027}
\BIBentrySTDinterwordspacing

\bibitem{gasser2018clusters}
O.~Gasser, Q.~Scheitle, P.~Foremski, Q.~Lone, M.~Korczynski, S.~D. Strowes, L.~Hendriks, and G.~Carle, ``Clusters in the expanse: Understanding and unbiasing ipv6 hitlists,'' in \emph{Proceedings of the 2018 Internet Measurement Conference}.\hskip 1em plus 0.5em minus 0.4em\relax New York, NY, USA: ACM, 2018.

\bibitem{gasser2022}
J.~Zirngibl, L.~Steger, P.~Sattler, O.~Gasser, and G.~Carle, ``Rusty clusters? dusting an ipv6 research foundation,'' in \emph{Proceedings of the 2022 Internet Measurement Conference}.\hskip 1em plus 0.5em minus 0.4em\relax New York, NY, USA: ACM, 2022.

\bibitem{10.1007/978-3-319-76481-8_10}
T.~Fiebig, K.~Borgolte, S.~Hao, C.~Kruegel, G.~Vigna, and A.~Feldmann, ``In rdns we trust: Revisiting a common data-source's reliability,'' in \emph{Passive and Active Measurement}, R.~Beverly, G.~Smaragdakis, and A.~Feldmann, Eds.\hskip 1em plus 0.5em minus 0.4em\relax Cham: Springer International Publishing, 2018, pp. 131--145.

\bibitem{6forest}
T.~Yang, Z.~Cai, B.~Hou, and T.~Zhou, ``6forest: An ensemble learning-based approach to target generation for internet-wide ipv6 scanning,'' in \emph{IEEE INFOCOM 2022 - IEEE Conference on Computer Communications}, 2022, pp. 1679--1688.

\bibitem{6graph}
T.~Yang, B.~Hou, Z.~Cai, K.~Wu, T.~Zhou, and C.~Wang, ``6graph: A graph-theoretic approach to address pattern mining for internet-wide ipv6 scanning,'' \emph{Computer Networks}, vol. 203, p. 108666, 2022.

\bibitem{6gan}
\BIBentryALTinterwordspacing
T.~Cui, G.~Gou, G.~Xiong, C.~Liu, P.~Fu, and Z.~Li, ``6gan: Ipv6 multi-pattern target generation via generative adversarial nets with reinforcement learning,'' in \emph{40th {IEEE} Conference on Computer Communications, {INFOCOM} 2021, Vancouver, BC, Canada, May 10-13, 2021}.\hskip 1em plus 0.5em minus 0.4em\relax {IEEE}, 2021, pp. 1--10. [Online]. Available: \url{https://doi.org/10.1109/INFOCOM42981.2021.9488912}
\BIBentrySTDinterwordspacing

\bibitem{6veclm}
\BIBentryALTinterwordspacing
T.~Cui, G.~Xiong, G.~Gou, J.~Shi, and W.~Xia, ``6veclm: Language modeling in vector space for ipv6 target generation,'' in \emph{Machine Learning and Knowledge Discovery in Databases: Applied Data Science Track: European Conference, ECML PKDD 2020, Ghent, Belgium, September 14–18, 2020, Proceedings, Part IV}.\hskip 1em plus 0.5em minus 0.4em\relax Berlin, Heidelberg: Springer-Verlag, 2020, p. 192–207. [Online]. Available: \url{https://doi.org/10.1007/978-3-030-67667-4-12}
\BIBentrySTDinterwordspacing

\bibitem{6hit}
\BIBentryALTinterwordspacing
B.~Hou, Z.~Cai, K.~Wu, J.~Su, and Y.~Xiong, ``6hit: {A} reinforcement learning-based approach to target generation for internet-wide ipv6 scanning,'' in \emph{40th {IEEE} Conference on Computer Communications, {INFOCOM} 2021, Vancouver, BC, Canada, May 10-13, 2021}.\hskip 1em plus 0.5em minus 0.4em\relax {IEEE}, 2021, pp. 1--10. [Online]. Available: \url{https://doi.org/10.1109/INFOCOM42981.2021.9488794}
\BIBentrySTDinterwordspacing

\bibitem{6gcvae}
T.~Cui, G.~Gou, and G.~Xiong, ``6gcvae: Gated convolutional variational autoencoder for ipv6 target generation,'' in \emph{Advances in Knowledge Discovery and Data Mining}, H.~W. Lauw, R.~C.-W. Wong, A.~Ntoulas, E.-P. Lim, S.-K. Ng, and S.~J. Pan, Eds.\hskip 1em plus 0.5em minus 0.4em\relax Cham: Springer International Publishing, 2020, pp. 609--622.

\bibitem{6tree}
\BIBentryALTinterwordspacing
Z.~Liu, Y.~Xiong, X.~Liu, W.~Xie, and P.~Zhu, ``6tree: Efficient dynamic discovery of active addresses in the ipv6 address space,'' \emph{Comput. Networks}, vol. 155, pp. 31--46, 2019. [Online]. Available: \url{https://doi.org/10.1016/j.comnet.2019.03.010}
\BIBentrySTDinterwordspacing

\bibitem{6scan}
\BIBentryALTinterwordspacing
B.~Hou, Z.~Cai, K.~Wu, T.~Yang, and T.~Zhou, ``6scan: {A} high-efficiency dynamic internet-wide ipv6 scanner with regional encoding,'' \emph{{IEEE/ACM} Trans. Netw.}, vol.~31, no.~4, pp. 1870--1885, 2023. [Online]. Available: \url{https://doi.org/10.1109/TNET.2023.3233953}
\BIBentrySTDinterwordspacing

\bibitem{addrminer}
\BIBentryALTinterwordspacing
G.~Song, J.~Yang, L.~He, Z.~Wang, G.~Li, C.~Duan, Y.~Liu, and Z.~Sun, ``Addrminer: {A} comprehensive global active ipv6 address discovery system,'' in \emph{2022 {USENIX} Annual Technical Conference, {USENIX} {ATC} 2022, Carlsbad, CA, USA, July 11-13, 2022}, J.~Schindler and N.~Zilberman, Eds.\hskip 1em plus 0.5em minus 0.4em\relax {USENIX} Association, 2022, pp. 309--326. [Online]. Available: \url{https://www.usenix.org/conference/atc22/presentation/song}
\BIBentrySTDinterwordspacing

\bibitem{hmap6}
\BIBentryALTinterwordspacing
B.~Hou, Z.~Cai, K.~Wu, T.~Yang, and T.~Zhou, ``Search in the expanse: Towards active and global ipv6 hitlists,'' in \emph{{IEEE} {INFOCOM} 2023 - {IEEE} Conference on Computer Communications, New York City, NY, USA, May 17-20, 2023}.\hskip 1em plus 0.5em minus 0.4em\relax {IEEE}, 2023, pp. 1--10. [Online]. Available: \url{https://doi.org/10.1109/INFOCOM53939.2023.10229089}
\BIBentrySTDinterwordspacing

\bibitem{DBLP:conf/sigcomm/RyeL23}
\BIBentryALTinterwordspacing
E.~C. Rye and D.~Levin, ``Ipv6 hitlists at scale: Be careful what you wish for,'' in \emph{Proceedings of the {ACM} {SIGCOMM} 2023 Conference, {ACM} {SIGCOMM} 2023, New York, NY, USA, 10-14 September 2023}, H.~Schulzrinne, V.~Misra, E.~Kohler, and D.~A. Maltz, Eds.\hskip 1em plus 0.5em minus 0.4em\relax {ACM}, 2023, pp. 904--916. [Online]. Available: \url{https://doi.org/10.1145/3603269.3604829}
\BIBentrySTDinterwordspacing

\bibitem{rfc7707}
\BIBentryALTinterwordspacing
I.~E. T.~F. (IETF), ``Network reconnaissance in ipv6 networks,'' 2016. [Online]. Available: \url{https://www.rfc-editor.org/rfc/rfc7707.html}
\BIBentrySTDinterwordspacing

\bibitem{ipv6_hitlist}
\BIBentryALTinterwordspacing
O.~Gasser, ``Ipv6 hitlist service,'' 2023. [Online]. Available: \url{https://ipv6hitlist.github.io/}
\BIBentrySTDinterwordspacing

\bibitem{det}
G.~Song, J.~Yang, Z.~Wang, L.~He, J.~Lin, L.~Pan, C.~Duan, and X.~Quan, ``Det: Enabling efficient probing of ipv6 active addresses,'' \emph{IEEE/ACM Transactions on Networking}, vol.~30, no.~4, pp. 1629--1643, 2022.

\bibitem{kingma2022autoencoding}
D.~P. Kingma and M.~Welling, ``Auto-encoding variational bayes,'' 2022.

\bibitem{NIPS2017_3f5ee243}
\BIBentryALTinterwordspacing
A.~Vaswani, N.~Shazeer, N.~Parmar, J.~Uszkoreit, L.~Jones, A.~N. Gomez, L.~u. Kaiser, and I.~Polosukhin, ``Attention is all you need,'' in \emph{Advances in Neural Information Processing Systems}, I.~Guyon, U.~V. Luxburg, S.~Bengio, H.~Wallach, R.~Fergus, S.~Vishwanathan, and R.~Garnett, Eds., vol.~30.\hskip 1em plus 0.5em minus 0.4em\relax Curran Associates, Inc., 2017. [Online]. Available: \url{https://proceedings.neurips.cc/paper-files/paper/2017/file/3f5ee243547dee91fbd053c1c4a845aa-Paper.pdf}
\BIBentrySTDinterwordspacing

\bibitem{10.1145/3422622}
\BIBentryALTinterwordspacing
I.~Goodfellow, J.~Pouget-Abadie, M.~Mirza, B.~Xu, D.~Warde-Farley, S.~Ozair, A.~Courville, and Y.~Bengio, ``Generative adversarial networks,'' \emph{Commun. ACM}, vol.~63, no.~11, p. 139–144, oct 2020. [Online]. Available: \url{https://doi.org/10.1145/3422622}
\BIBentrySTDinterwordspacing

\bibitem{addrminer_2.0}
\BIBentryALTinterwordspacing
G.~Song, ``Addrminer-v2.0,'' 2023. [Online]. Available: \url{https://github.com/AddrMiner/AddrMiner-v2.0}
\BIBentrySTDinterwordspacing

\bibitem{RIPEstat}
\BIBentryALTinterwordspacing
RIPE, ``Ripestat data api,'' 2023. [Online]. Available: \url{https://stat.ripe.net/docs/02.data-api/}
\BIBentrySTDinterwordspacing

\bibitem{rfc6177}
\BIBentryALTinterwordspacing
I.~E. T.~F. (IETF), ``Ipv6 address assignment to end sites,'' 2011. [Online]. Available: \url{https://datatracker.ietf.org/doc/html/rfc6177}
\BIBentrySTDinterwordspacing

\bibitem{auer2002finite}
P.~Auer, N.~Cesa-Bianchi, and P.~Fischer, ``Finite-time analysis of the multiarmed bandit problem,'' \emph{Machine learning}, vol.~47, pp. 235--256, 2002.

\bibitem{ethic1}
\BIBentryALTinterwordspacing
C.~Partridge and M.~Allman, ``Ethical considerations in network measurement papers,'' \emph{Commun. ACM}, vol.~59, no.~10, p. 58–64, sep 2016. [Online]. Available: \url{https://doi.org/10.1145/2896816}
\BIBentrySTDinterwordspacing

\bibitem{ethic2}
D.~Dittrich, E.~Kenneally, and M.~Bailey, ``Applying ethical principles to information and communication technology research: A companion to the menlo report,'' \emph{Available at SSRN 2342036}, 2013.

\bibitem{SLAAC}
\BIBentryALTinterwordspacing
N.~W. Group, ``Ipv6 stateless address autoconfiguration,'' 2007. [Online]. Available: \url{https://datatracker.ietf.org/doc/html/rfc4862}
\BIBentrySTDinterwordspacing

\end{thebibliography}
